\documentclass[aps,pre,twocolumn,groupedaddress,showpacs]{revtex4}
\usepackage{amsmath}
\usepackage{amssymb}
\usepackage{graphicx}
\usepackage{epsfig}
\usepackage{bbold}
\usepackage{subfigure}
\newcommand{\be}{\begin{equation}}
\newcommand{\ee}{\end{equation}}
\newcommand{\bea}{\begin{eqnarray}}
\newcommand{\eea}{\end{eqnarray}}

\newcommand{\kt}{k_{\rm B}T}

\newcommand{\ve}{\varepsilon}

\def\lsim{\mathrel{\rlap{\lower4pt\hbox{$\sim$}}
    \raise1pt\hbox{$<$}}}                
\def\gsim{\mathrel{\rlap{\lower4pt\hbox{$\sim$}}
    \raise1pt\hbox{$>$}}}                
    
\begin{document}


\title{Self-assembling multiblock amphiphiles:\\ molecular design, supramolecular structure, and mechanical properties}



\author{Hamed Mortazavi and Cornelis Storm}
\affiliation{Department of Applied Physics and
Institute for Complex Molecular Systems, 
Eindhoven University of Technology, 
P.O. Box 513, 5600 MB Eindhoven, The Netherlands}

\date{\today}

\begin{abstract}
We perform off-lattice, canonical ensemble molecular dynamics simulations of the self-assembly of long segmented copolymers consisting of alternating, tunably attractive and hydrophobic {\em binder} domains, connected by hydrophilic {\em linker} chains whose length may be separately controlled. In such systems, the molecular design of the molecule directly determines the balance between energetic and entropic tendencies. We determine the structural phase diagram of this system, which shows collapsed states (dominated by the attractive linkers' energies), swollen states (dominated by the random coil linkers' entropies) as well as intermediate network hydrogel phases, where the long molecules exhibit partial collapse to a {\em single molecule network} state. We present an analysis of the connectivity and spatial structure of this network phase, and relate its basic topology to mechanical properties, using a modified rubber elasticity model. We find that it is possible to optimize the mechanical performance by an appropriate choice of molecular design, which may point the way to novel synthetics that make optimal mechanical use of constituent polymers.  
\end{abstract}
 

\maketitle 

\section{Introduction}

Biological materials consist mostly of networks of filamentous fibers. A prime example is the cytoskeleton, which - among its many other functions - governs mechanical stability, the response to stress, and the motility of the cell. Its mechanical properties, particularly in the non-linear regime are uniquely linked to its molecular architecture: an open meshworks of interconnected semi-flexible fibers \cite{BauschKroy,ChaseReview}. Mimicking these mechanical properties in a synthetic material would offer important new possibilities both for cell research and health technologies such as drug delivery or cell growth, as well as for novel bio-insipred and biobased performance materials. In this paper, we investigate a self-assembly route to recreating the topology of such materials in synthetic associative polymers, and ask the question how these should be designed in order to spontaneously produce networks that present optimal mechanical performance. To this end we study, numerically, a self-assembling system of block copolymers that forms an effective cross-linked network. 

\begin{figure}[ht]\centering
\includegraphics[width=.5\textwidth]{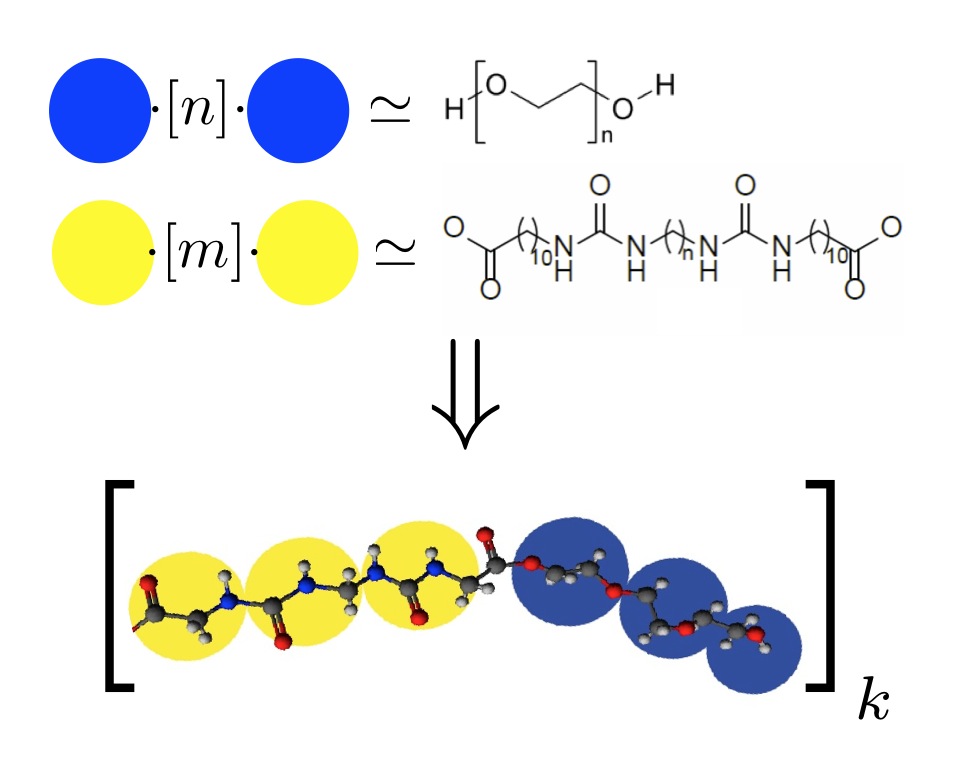}
\caption{We simulate long copolymers consisting of alternating binder (yellow) and linker (blue) blocks. The binder blocks consist of repeating PEG units, the linker blocks are bisurea motifs. In synthesis, the length of both these blocks is separately tuneable. In our simulations, we coarse grain these molecules into beads representing many atoms at once. The molecular design is fixed by specifying values for $n$, $m$ and $k$.}    
\label{fig1}
\end{figure}

The self assembling system we focus on here is a multiblock amphiphile, consisting of many regular repeats of alternating hydrophilic and hydrophobic blocks. Previous work, which we will briefly review in a moment, has identified such molecules as quite promising candidates for precise regulation of self-organized architecture, but so far experimental realizations have largely been lacking. Polysaccharides with hydrophobic substitutions - methylcellulose, in particular - possess the required alternatingly hydrophilic/hydrophobic backbone structure, but in more or less randomly distributed blocks \cite{sarkar}. Recently, however, a novel molecule was synthesized and studied \cite{pawar2012injectable} that does allow precise control over the spatial arrangement of the alternating blocks, and in fact also over the attractiveness of the individual hydrophobic blocks. Its basic building blocks (shown in Fig. \ref{fig1}) are macromolecules consisting of alternating binder (attractive) and linker blocks, which are, respectively, bis-urea hydrophobic motifs (essentially identical to those discussed in \cite{Koenigs}), and polyethylene glycol (PEG) hydrophilic chains \cite{pawar2012injectable}. The objective of this work is to provide an initial survey, based upon Molecular Dynamics simulation and elementary rubber elastic theory, of the expected microstructure of the self-assembled states of this molecule, as well as the mechanical properties of its gel/network phase.

In our simulations, we coarse grain many repeat units into beads that interact via standard hard-core repulsive and - for the binder blocks - attractive potentials. 

In aqeous solvent the amphiphilic design of these molecules leads to aggregation and causes the long molecules to partially collapse into clusters that are connected via one or more links. Clearly, if the attraction is too weak the chains will adopt random coil configurations, while if the attraction is too large the molecules will collapse on themselves forming compact globular aggregates. In between, as we will show, there is an intermediate regime where the system spontaneously aggregates into a network phase of connected clusters linked to each other. Here we present simulations in which we control the linker length and the interaction potential between the binders to tune the relative importance of energy and entropy in this self-assembly process, and characterize the resultant network in terms of cluster size and connectivity. We summarize these findings in a phase diagram, identifying the experimentally most promising regime for recovering biopolymer network-like architectures in self-assembling copolymers. We then analyze the mechanical properties of these networks, first in the context of a rubber-elastic theory for networks of variable functionality, then in direct numerical simulations. We are not the first to consider models of self assembling multiblock copolymers: Rooted in extensive work on the assembly and rheology of amphiphilic block copolymers (see, e.g., \cite{Lindman} and references therein) more recent work has also explicitly considered the multiblock design. Early scaling theory by Halperin \cite{Halperin} anticipated the emergence of flowerlike micellar structures, either single or multiple such structures connected to each other in the manner of pearls on a string, depending on the quality of the solvent. Subsequent efforts \cite{Zhang2003, Kawata2007} refined the conditions for formation and stability of the flowerlike micelles. Later numerical work \cite{Tanaka1, Tanaka2} confirmed that indeed such structures may form in dilute systems. In \cite{Hugo2009,HugoSM2011}, the Monte Carlo (MC) technique on a lattice model was applied on the system to establish an equilibrium phase diagram, which features characteristic phases of isloated micelles, connected micelles, but also laminar and tubular phases. Similar results were reported in \cite{Tanaka1,CookeWilliams} in yet more MC simulations. More recently, the first dynamical simulations (Brownian Dynamics, BD) were presented, varying systematically the solvent quality. Again, the familiar structural phases were reported \cite{Sun2013}. The picture that emerges is a very rich self-assembling system, which - provided one has some synthetic control over the moelcular architecture - provides a direct control over mesoscopic organization of a hydrogel. A question that has received very little attention, so far, has been the implications of such structure for mechanical performance. Our ultimate ambition is to simulate dynamical rheology for these networks, in conjunction with the self-assembly. We focus on those structural aspects of the hydrogel phase that govern mechanical response. This response is analyzed first in the context of simple rubber elastic models, and then directly, using a computational implementation of the oscillatory rheology protocol on the networks presented here. While - aside from the polysaccharides mentioned above - there are few experimental realizations to compare to, the molecular design we coarse grain here was, in fact, studied in previous experimental work \cite{pawar2012injectable}, which revealed a tendency to form strong hydrogels, whose rheological properties make them uniquely suited as injectable substrates for drug delivery. However, despite the detailed AFM and SAXS analysis of the self-assembled hydrogels, the question of the precise molecular structure of the gels remained unanswered. In this paper, we use molecular dynamical simulations to address this question, which enables us to determine a relation between molecular design, gel structure, and mechanical properties that breaks ground for further rational design in these materials. 

This paper is organized as follows: Section \ref{sec2} presents the experimental model system, the modeling environment and approach we have adopted, and the relevant interaction potential we employ. In Section \ref{sec3} the simulation protocol in coarse-graining the model system is presented. In Section \ref{sec4}, we present data on the phase behavior resulting from the self-assembly process. Section \ref{sec5} analyzes the topology of the network phase and presents results on the size and distribution of the connectivity of clusters. In the remainder of the paper we investigate the mechanical response of these networks, in Section \ref{sec6}, we recall and apply a modified rubber elastic model, in Section \ref{sec7} we measure the rheological response of such networks numerically, using a computational bulk rheology approach. This section is followed by qualitative result section. We then relate these findings to experimental observations. We present our conclusions for the relation between molecular design, supramolecular network structure and the mechanical properties in our system. 

\section{Model system and coarse grained MD setup}
\label{sec2}

As indicated in Fig. \ref{fig1}, the multiblock copolymer we study here is a polymer consisting of repeating diblock units. Its structure may be denoted as [[PEG]$_n$[bisurea$_m$]]$_k$; $n$, $m$ and $k$ are the repeat numbers of the PEG molecule, the bisurea motif, and the resultant diblock unit respectively. In typical experimental settings, $n\sim 10^2, m \sim 5, k \sim 10$ \cite{pawar2012injectable}. The PEG chain is hydrophilic and highly flexible which results in a tendency to swell in water due to entropy maximization. The bis-urea block, in contrast, is stiff (nonswellable) and hydrophobic, resulting in a generic tendency to aggregate in water. This tendency is further enhanced in by the ability of the ureas to hydrogen bond, leading to an increased stability of their aggregates. Thus, the ultimate behavior of this long copolymer is determined by conflicting tendencies imparted by binders and linkers; depending on the dominating feature the system will lean towards collapsed (binder dominated) states or swollen (linker dominated) states. Interestingly, the geometry of the molecule may be used to directy affect this balance between energy and entropy. The longer the linker chain ($n$), the larger the entropic contribution; the longer the bisurea block ($m$) the stronger the hydrophobicity/hydrogen bond-driven tendency to collapse.\\

 \begin{figure}[ht]\centering
  \includegraphics[width=.5\textwidth]{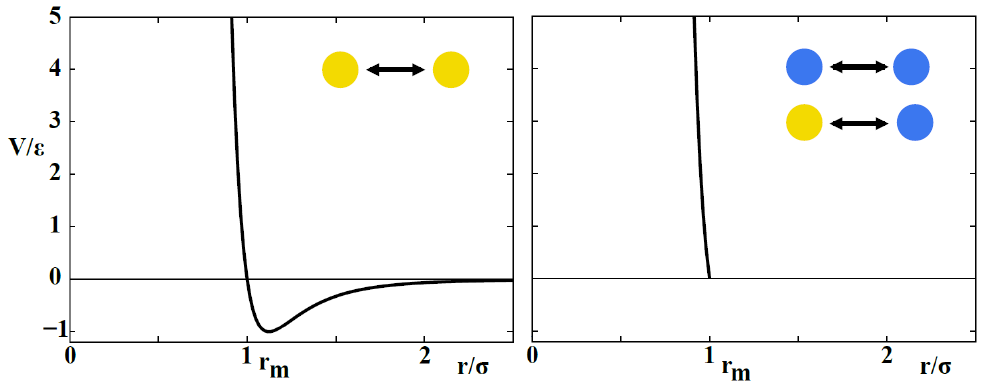}
  \caption{Graph of the Lennard-Jones (LJ) potentials between binder beads (left). Linker beads, interacting with either other linker beads or binder beads do so with a LJ potential truncated at $r=\sigma$ to approximate hard-core repulsion between beads(right). }    
   \label{fig:forcfield}
  \end{figure}

In our simulations, we are initially interested in the generic features of this class of systems. We therefore model a copolymer chain as a sequence of permanently attached beads of fixed (and equal) radius $\sigma$. There are two distinct types of beads, and they are distinguished by their mutual interactions (see Fig. \ref{fig:forcfield}). All beads repel strongly at distances shorter than $\sigma$, to enforce hard-core repulsion and non-crossing of the copolymer (we address this in more detail below). Binders mutually attract at distances larger than $\sigma$, whereas linkers experience no interactions, either with other linkers or binders, beyond their hard-core radius. As we are initially interested in open, meshworked phases we shall use as our two main control parameters in this system the number of binder beads $N_b$, the number of linker beads $N_l$, and the strength of the LJ attraction as measured by the LJ well depth $\ve$. As the experimental molecules have either $m=4$ or $m=6$, the physical size of the binder region does not vary much, which is why we choose to fix $N_b=3$, reducing the number of free parameters. Simulated coarse grained molecules are denoted as B$(N_b)$L$(N_l)$, so that, for instance $B3L24$ denotes a molecule whose repeating diblock motif consists of 3 binder beads and 24 linker beads. For all simulations reported here, we use 500 repeats of this diblock, so that each molecule in each simulation contains 1500 binder beads.\\
  \begin{figure}[ht]\centering
  \includegraphics[width=.5\textwidth]{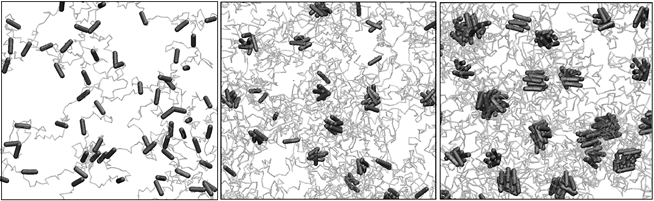}
  \caption{Evolution of the self-assembly of a network during a typical MD run in the network phase. Starting from a random arrangement, the macromolecule gradually collapses on itself to form clusters of a well defined size, connected by one or more linker chains. Note, that these images are still of single (albeit very large) molecules.}    
   \label{fig:snap}
  \end{figure}

As was previously observed in lattice MC and BD simulations \cite{Hugo2009,HugoSM2011,Sun2013}, the equilibrium phase diagram of such systems is already very rich. At low values of $\ve$, or very large linker regions, the entropy dominates and swollen phases are expected. For very high values of the LJ attraction, or for short linkers, energy dominates and complete collapse (similar to a polymer in poor solvent) is observed. In the intermediate regime, chains of micelles are reported, as are nonspherical aggregates. The objective of the present paper is to establish whether these structures are also present in off-lattice MD simulations. As will show, most do indeed feature in our simulations, although a large region of the dynamical phase space is occupied by meshlike structures which are best termed flowerlike micellar networks, which appear to be dynamically arrested intermediates that are unable to fully equilibrate to a single collapsed state, see Fig. \ref{fig:snap}. As these flowerlike micellar networks most closely resemble the biopolymer gels whose topology we aim to recreate, we characterize the connectivity structure and relate it to mechanical response using modified rubber elastic theory. 
  
Our simulations are carried out using the LAMMPS molecular dynamics package \cite{plimpton1995fast}. We represent the long copolymers using a bead-spring model, under fixed boundary conditions, with two distinct types of beads to represent binder and linker segments.  Adjacent beads interact via a harmonic bond potential of the form $U_{bond} (r) = k_b (r - l_b)^ 2$, and a harmonic bending potential $U_{angle}(\theta) = k _a (\theta - \pi)^2$, is applied to each set of three neighboring binder beads (where $r$ is the distance between the centers of mass of pairs of beads, $l_b = 1.2 \,\sigma$ is the equilibrium bond length, $\sigma$ the size (diameter) of the beads and $\theta$ the angle formed by the two bonds that connect the middle bead to its two adjacent beads). $k_b$ and $k_a$ are the stretching and bending stiffnesses, which are both fixed at fairly high values:  $k_b = 200 \, k_BT / \sigma^2$ and $k_a = 400 \, k_BT$ , to enforce inextensibility of the backbone, and inflexibility of the binder region during the simulations. The bending stiffness is set to zero for linker beads, so that the linkers are represented as flexible polymers of molecular weight proportional to $N_l$. All binder beads that are not first or second nearest neighbors (i.e., those that are part of different binder domains) interact through a full Lennard-Jones (LJ) potential, whose attractive strength measured by $\ve$:

\begin{equation}
\label{LJ}
U(r)=4\,\ve\left[\left( \frac{\sigma}{r}\right)^{12} - \left(\frac{\sigma}{r}\right)^{6}\right]	
\end{equation}

The interactions involving linker beads (i.e, linker-linker and binder-linker) are purely repulsive, and are modeled with a truncated LJ potential, (see Fig. \ref{fig:forcfield})
\begin{equation}\label{repLJ}
U(r)=\begin{cases}4\,\ve_{\rm rep}\left[\left( \frac{\sigma}{r}\right)^{12} - \left(\frac{\sigma}{r}\right)^{6}\right]	 &\text{if $r \leq \sigma $}\\
		0&\text{if $r> \sigma$,}\end{cases}
\end{equation}

\begin{figure}[ht]\centering
  \includegraphics[width=.5\textwidth]{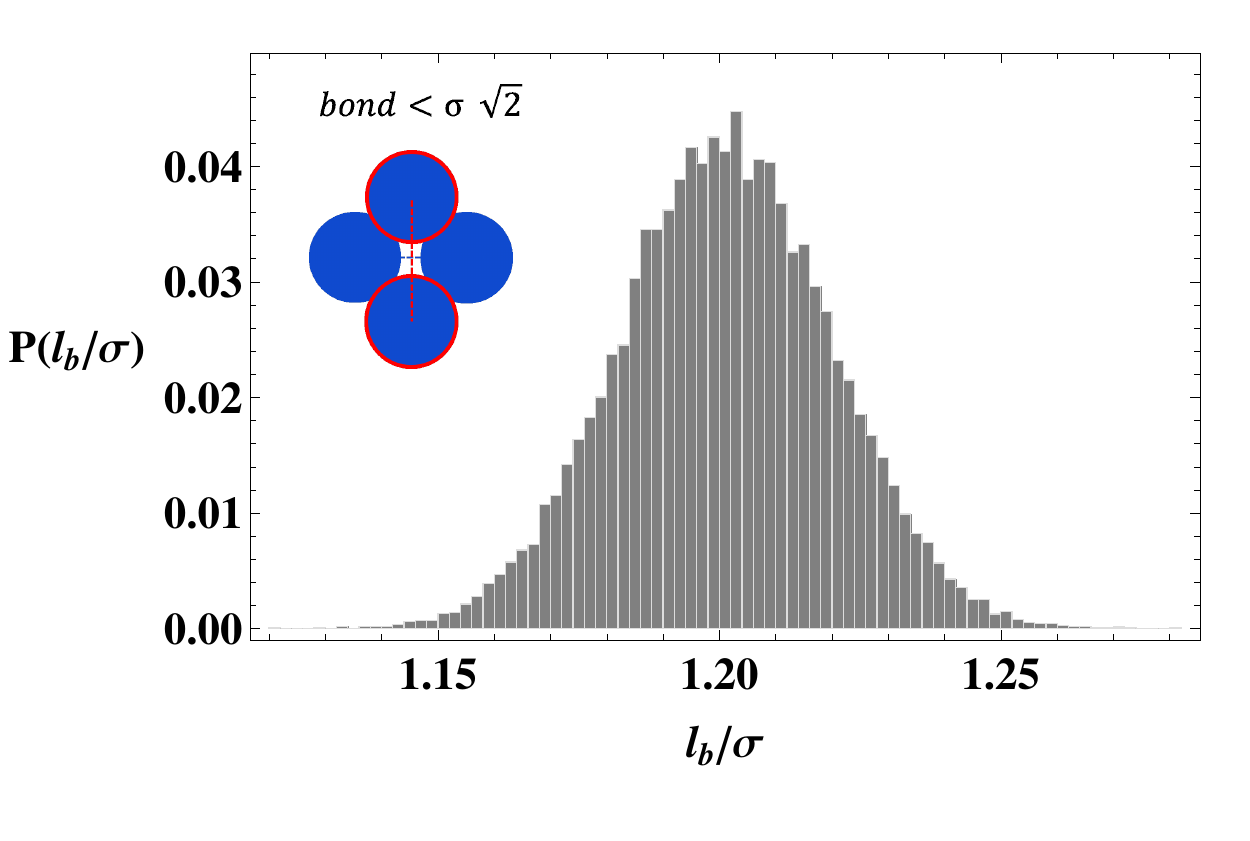}
  \caption{Histogram of bond length distribution during one run. The spacing between two beads is never sufficiently large to allow other beads through, and hence the chain cannot self cross during the dynamics.}    
   \label{fig:bond}
  \end{figure}

In the latter equation, $\varepsilon_{\rm rep}$ measures the strength of the repulsion, which in our simulations is set equal to $0.01 \,\kt$; $r$ is still the center-to-center distance between the beads. The repulsive potential is used to ensure that the polymer obeys self-avoidance: We geometrically ensure that the chain cannot cross itself by setting the bead-bead distance to 1.2$\sigma$, and we truncate the LJ potential at $r=\sigma$. The repulsive LJ potential takes care of the self avoidance of all beads, and to prohibit self-crossing of the polymer we choose an equilibrium bond length $l_a$ of 1.2$\sigma$. This leaves a gap with an equilibrium size of 0.2$\sigma$ between neighbouring beads, and as one may see in Fig. \ref{fig:bond}, the bond length never fluctuates to lengths sufficient to leave a space through which another segment of polymer may pass. 

To verify that indeed, this reproduces correct polymer scaling, we turn off the attractive interactions (setting $\ve_{\rm rep}$, as well as the bond-bending term) and record the radius of gyration as a function of the length of the polymer. As one may see in Fig. \ref{fig:self avoidance gen}, we recover the correct self-avoiding scaling $R_g \sim L^\nu$  with a Flory exponent $\nu$ of about 0.6. This scaling is unaffected by the actual size of beads. The repulsive LJ potential \ref{repLJ} exists between all beads, its value affects the effective radius of a bead. If we choose it to be too small, the beads become 'soft' and may overlap slightly. As one may see from Fig. \ref{fig:self avoidance gen}, the exponent $\nu$ rises quickly from $0.5$ (the ideal chain value) to its self-avoiding value upon $0.6$ for increasing values of $\ve_{\rm rep}$. We fix $\ve_{\rm rep}$ at 0.01 because at this value, we first see (within our numerical accuracy) the correct exponent of $\nu=0.6$.

\section{Simulation Protocol}
\label{sec3}
In our LAMMPS simulations, each simulation is repeated from random initial conditions for $5$ different systems, which are then averaged over. Our protocol is as follows: We begin each simulation by an equilibration process starting from a random initial geometry. To capture the effects of different attractive strengths (representing the different cohesive energies, which rises with $m$ in the molecule) and different linker lengths, we simulate  twelve types of coarse grained molecules: B3L3, B3L4, B3L5, B3L6, B3L9, B3L12, B3L18, B3L24, B3L45, B3L75, B3L120, B3L150, and each of these is simulated for twelve values of $\ve$. After an initial randomization, we turn on the attractive interactions and watch the system evolve.  

 \begin{figure}[ht]\centering
  \includegraphics[width=.5\textwidth]{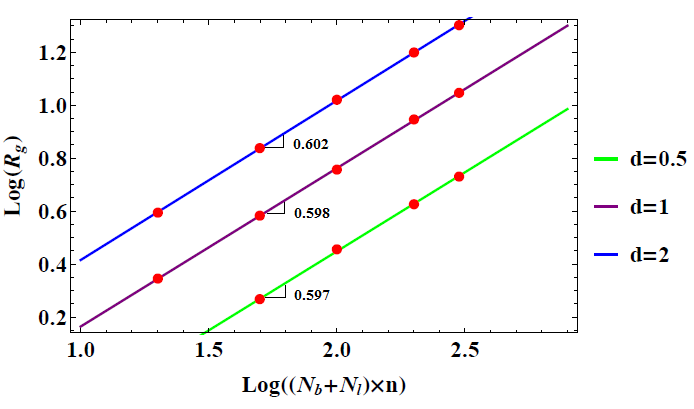}
  \caption{$R_g$ scaling for contour lengths of $20,50,100,200$ and $300$ beads. The attractive potentials are switched off. The Flory exponent for beads of diameters of $d=0.5, 1$ and $2$ is in agreement with theoretical prediction.}    
   \label{fig:self avoidance}
  \end{figure}

 \begin{figure}[ht]\centering
  \includegraphics[width=.5\textwidth]{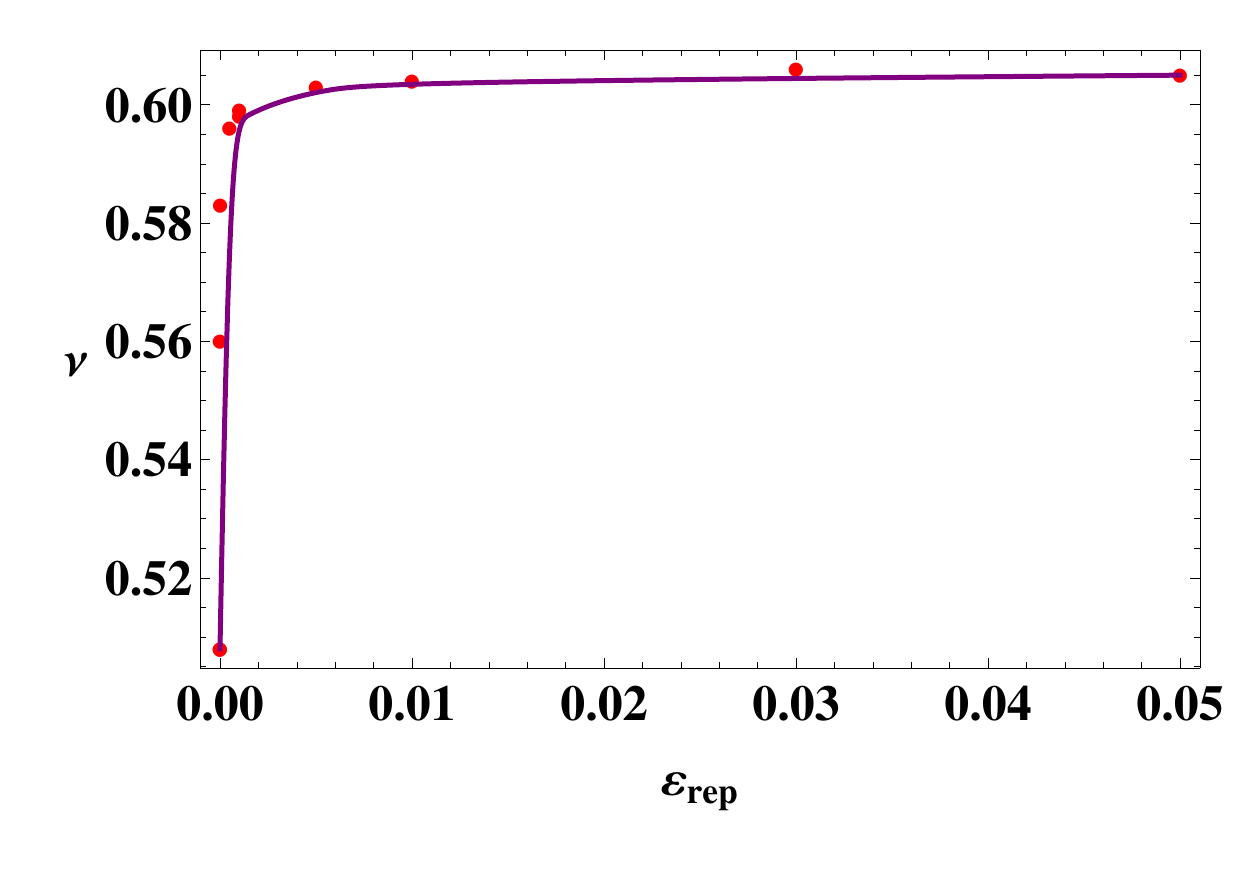}
  \caption{Scaling exponent $\nu$ for different $\varepsilon_{rep}$. We have chosen $\varepsilon_{rep}=0.01$ in all our simulations which as this is the value where correct real chain scaling is first observed.}    
   \label{fig:self avoidance gen}
  \end{figure}

\section{Phase Diagram}
\label{sec4}
\begin{figure}[ht]\centering
  \includegraphics[width=.5\textwidth]{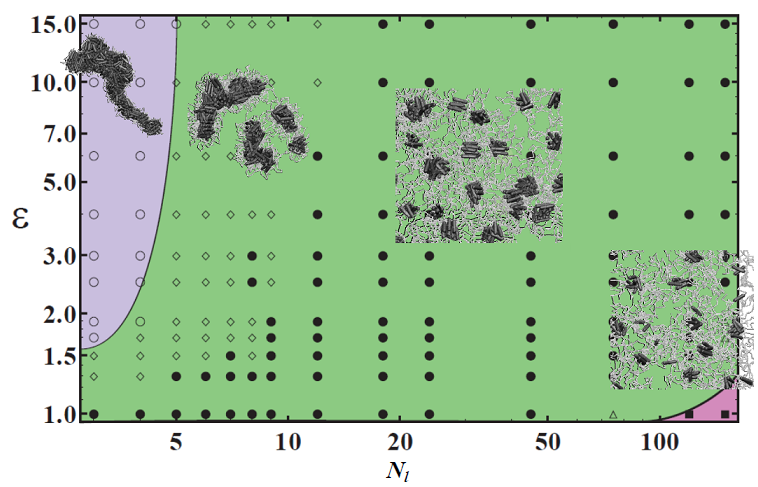}
  \caption{The phase diagram of the self-assembled states of the macromolecule is divided into three distinct phases: large single clusters (violet), single molecule networks (green) and random coils (purple). The network phase is also loosely divided into three sub-phases: One with a broad distribution of cluster sizes ($\diamondsuit$), ane with much tighter distribution of cluster sizes ($\bullet$), and one with a combination of clusters and isolated single beads ($\bigtriangleup$). }    
   \label{fig:phase1}
  \end{figure}

What happens next obviously depends on the strength of the attractive interactions, and the geometry of the molecule. For low values of $\ve$,
much like swollen polymers in a good solvent, we see random coil configurations, determined by the entropy of the linkers. Upon increasing $\varepsilon$, the chains start to self-assemble and form clusters. We see the size of clusters grow with increasing $\varepsilon$ but their growth rate is limited by the entropy of the linkers. For short linkers, collapse to a single cluster at high $\ve$ values is observed. For even longer linkers, however, the average cluster size remains limited since the entropic penalty for packing the linkers into the corona of the aggregate rises strongly with their length. The intermediate regime is where we see a network of flowerlike micellar cores (small aggregates of binders), linked to other such clusters by one or more linker chains - see Fig. \ref{fig:phase1}. This phase diagram is in good agreement with a similar result reported in earlier lattice Monte Carlo simulations, though we obviously cannot reproduce the lattice-induced layered aggregate phase reported there. \\

A second, more striking difference is that the network phase we see appears to be more prevalent in our MD simulations than in the earlier MC work \cite{HugoSM2011,Hugo2009,Sun2013}. We attribute this to kinetic trapping of the structures: while truly long relaxation might recover collapsed phases as true thermal equilibria, on the timescales that we are able to access in our MD sims the system tends to arrest in long lived, but possibly metastable, stationary states. In the following, we characterize the topology in this network phase.

\section{Network structure analysis}
\label{sec5}
Clearly, the region in the phase diagram that most closely resembles that of a crosslinked biopolymer mesh is the network phase. As is known from extensive previous work \cite{Wilhelm2003,Gardel2004,Gardel2003,Storm2005,Huisman2007,ChaseReview,BauschKroy}, the topology of such a network (connectivity, crosslinker density) greatly influence the mechanical response. We will quantify the connectivity and cluster size of the networks that form in the intermediate regime, as these parameters are expected to directly translate into effective parameters for the crosslinked network: We identify each cluster with a network node, whose cohesive energy is determined by the number of binders it contains, and we count the number of connecting links between such network nodes to determine the distribution of functionalities in the network.

  \begin{figure}[ht]\centering
  \includegraphics[width=.45\textwidth]{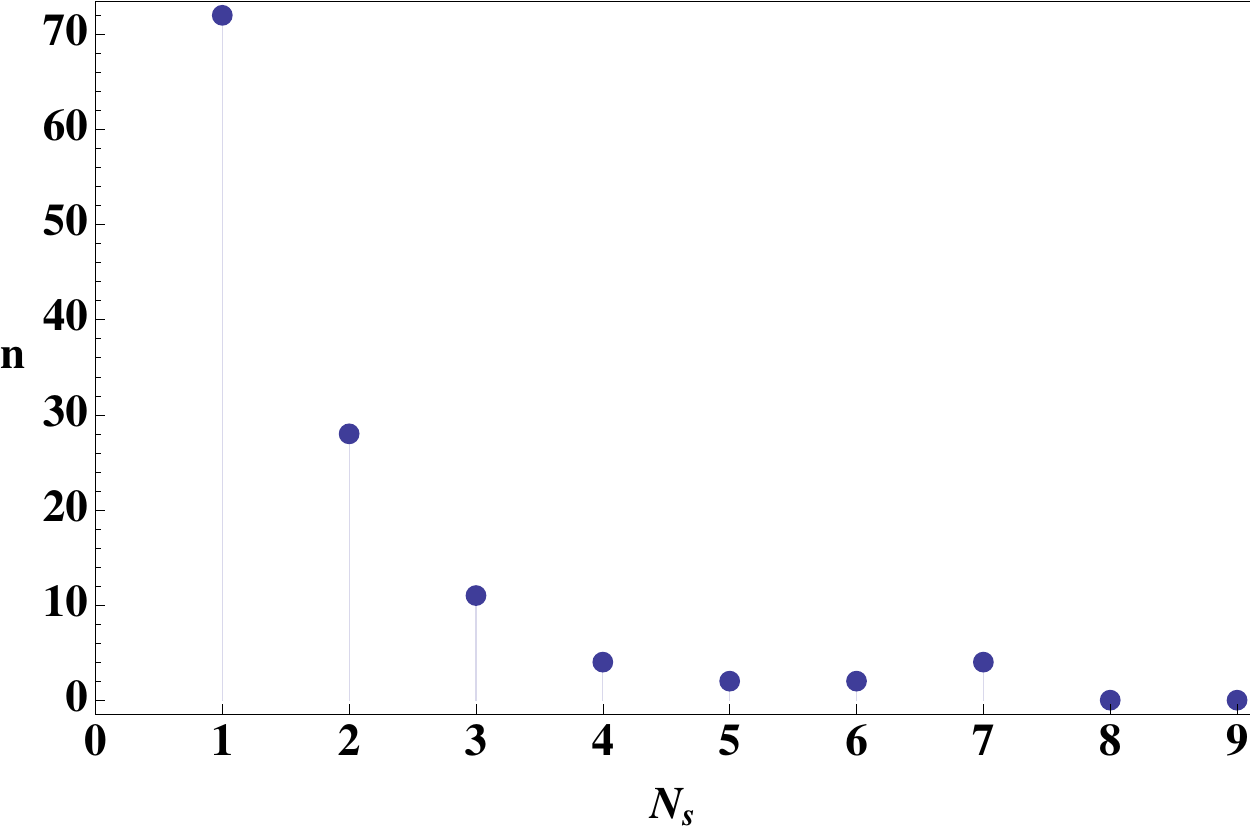}
  \caption{Frequency distribution of the step length $N_s$ for B3L7, at $\ve=1.5$. The rapid decay justifies the approximation we will make, which is that all linkers are one step linkers.}    
   \label{fig:functionality}
  \end{figure}

Our method to identify the clusters is as follows: After equilibration, we export the MD trajectories of all beads. To identify individual clusters, we sort all binders that are within $d_{\rm max}$ from each other in a group. $d_{\rm max}$ is the mean value of the longest distance between two center of mass $(3.4 \sigma)$  (the heart-to-heart distance between the central beads of two aligned binder blocks) and the shortest such distance $(\sigma)$ (i.e., the heart-to-heart distance between the central beads of two parallel binder blocks). The average cluster size is then an average over the distribution of number of binders in each group. This computation is averaged over five separate runs. We also calculate the {\em functionality} $f$ of each cluster, defined as the number of linker chains that bridge to other clusters emanating from one particular cluster. In our simulations we count all of the chains that connect two separate clusters together. However, the length of these bridges can be one or more times the linker length, and the length of the linker determines its effective elasticity. To keep track of this, we also record  the {\em step length} $N_s$ of bridges, denoting the number of linkers $n$ that connect one cluster to the next. We find, however, that the distribution of the step length decays rapidly with $N_s$, suggesting that the initial approximation that most bridges are one-step is justified - see Fig. \ref{fig:functionality}.

\begin{figure}[ht]\centering
  \includegraphics[width=.5\textwidth]{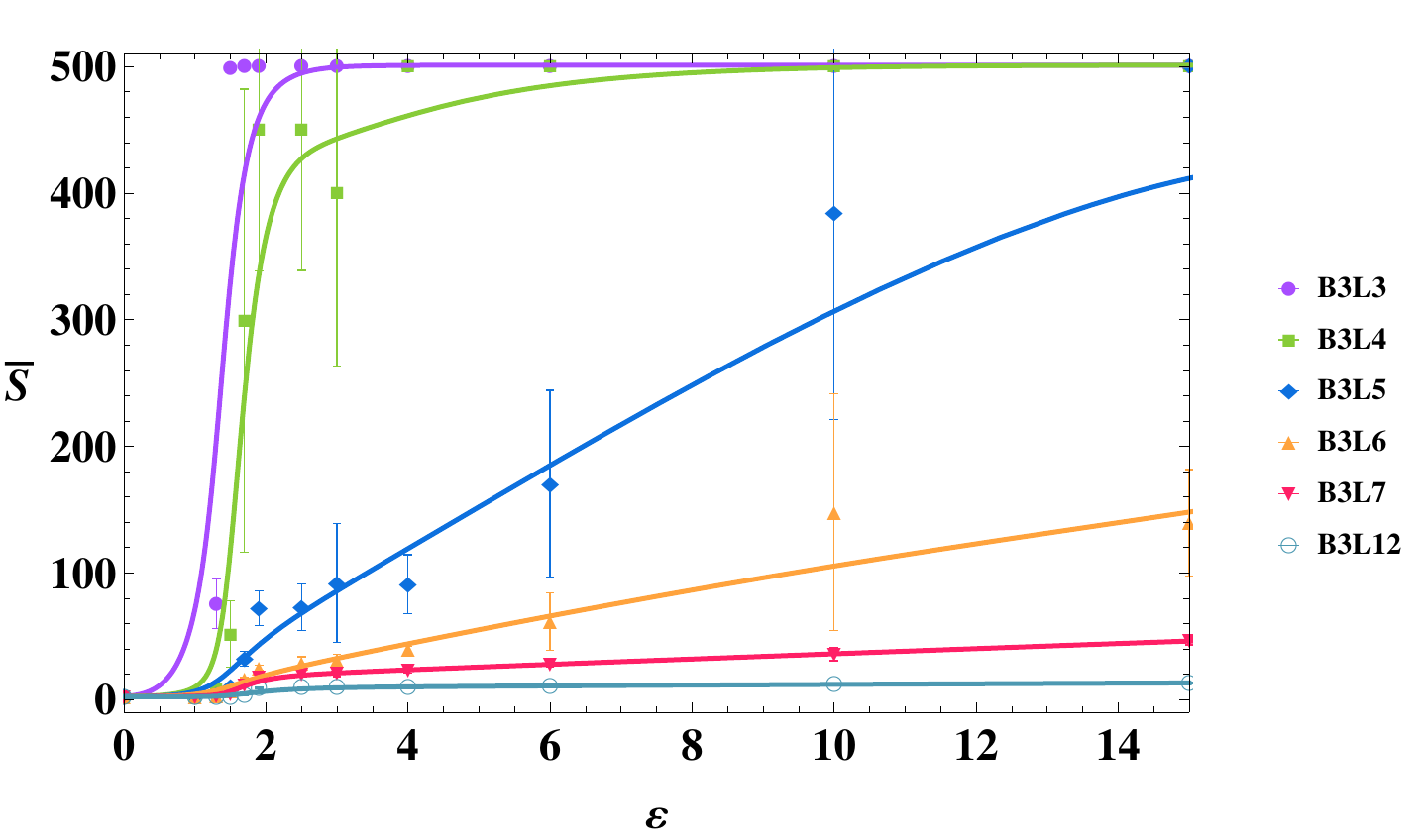}
  \caption{Average cluster size vs $\varepsilon$. Networks with large $\varepsilon$ tend to form bigger clusters. This tendency is more pronounced in molecules with shorter linkers.}     
  \label{fig:CvsE_general}

\includegraphics[width=.5\textwidth]{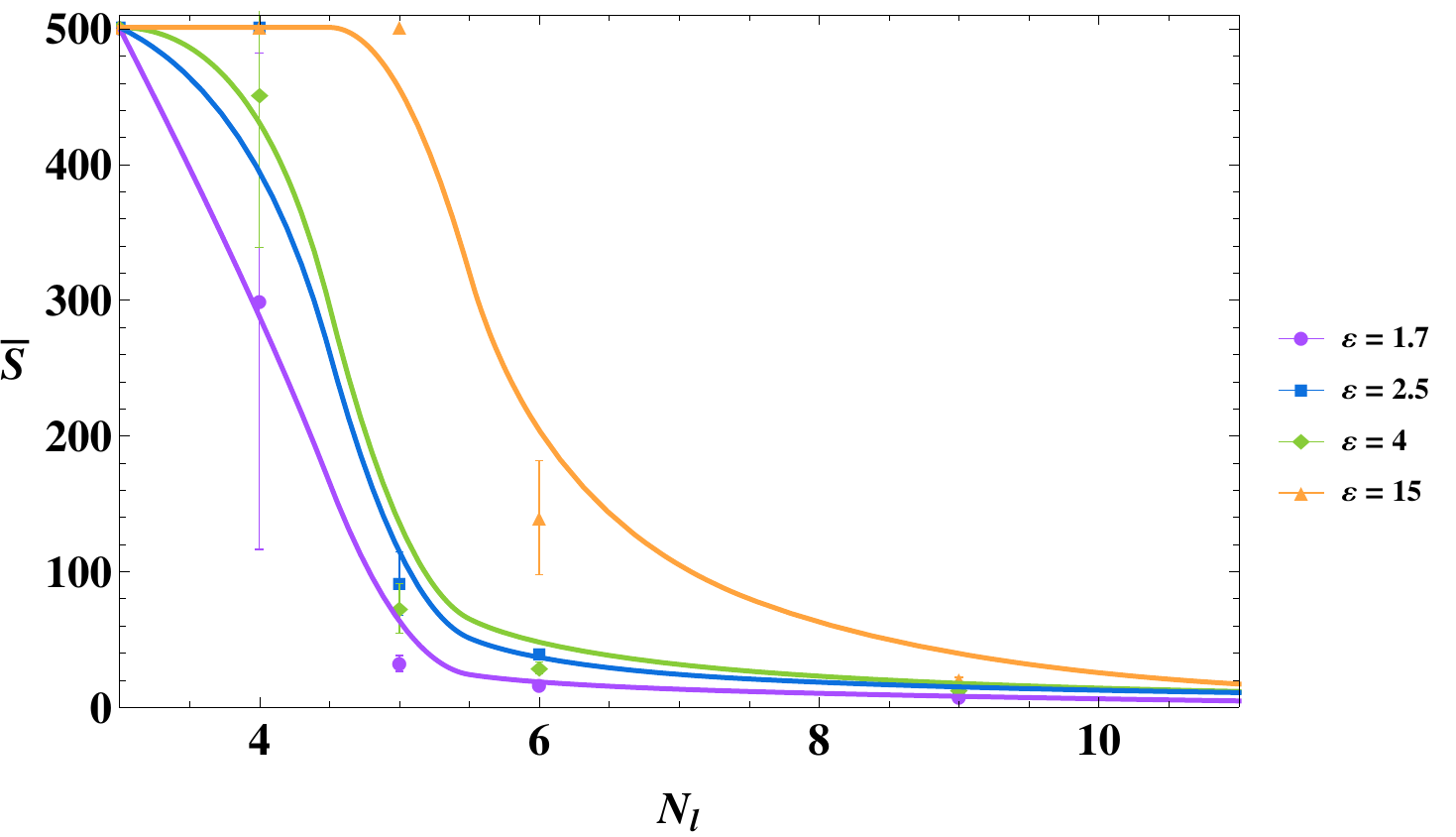}
  \caption{Average cluster size vs linker length. Increasing the proportion of linker beads increases the contribution of the entropy of the linker to the overall free energy, favoring smaller clusters.} \label{fig:CvsLL_general} 
  \end{figure}

As Fig. \ref{fig:CvsE_general} shows, the cluster size increases with increasing attractive LJ strength $\ve$. Basically, we are increasing the importance of energetic over entropic effects. The upper bound of 500 on the cluster size is determined by our choice of system - as stated we have 500 total binder motifs in each simulation (i.e., 1500 binder beads). A system that reaches a cluster size of 500 is therefore fully collapsed. For very low values of $\ve$, the cluster size is 1 which means no binder is connected to any other binder, and the system is swollen. The cluster size measure is thus able to distinguish between the extremes of swollen random coil phase and collapsed state. For intermediate values of $\ve$, the system settles into the network phase (see also Fig. \ref{fig:CvsE_general}) and exhibits a finite cluster size, below 500 but significantly greater than one.

The other way in which we can alter the balance between energy and entropy is by increasing the linker length. The longer the highly flexible linker chains become, the more configurational entropy they possess. This is indeed borne out by Fig. \ref{fig:CvsLL_general}, which shows that for short linkers, we generally see full collapse but that for larger values of $N_l$, the system swells and - at a rate that depends on $\ve$ - reverts to the fully random swollen state. 

Clearly, the combination of $\ve$ and $N_l$ determines the size of the clusters, as well as the connectivity between them. We may now ask what the projected mechanical properties of the resultant network phase will be. These too will depend on the structure of the network: in a dilute system, such as the ones we consider here, both the random coil state and the collapsed state are expected to have poor mechanical performance, if not complete lack of rigidity - below the overlap concentration, distinct molecules will not entangle to any significant degree in the random coil phase, and most certainly will not do so in the collapsed state. In the intermediate, network phase different molecules will perticipate in a single, connected network of flowerlike micelles. In the following, we estimate the mechanical modulus of such a network based on the structural features we have just discussed. 

\section{the mechanical response}
\label{sec6}

\begin{figure}[ht]\centering
  \includegraphics[width=.5\textwidth]{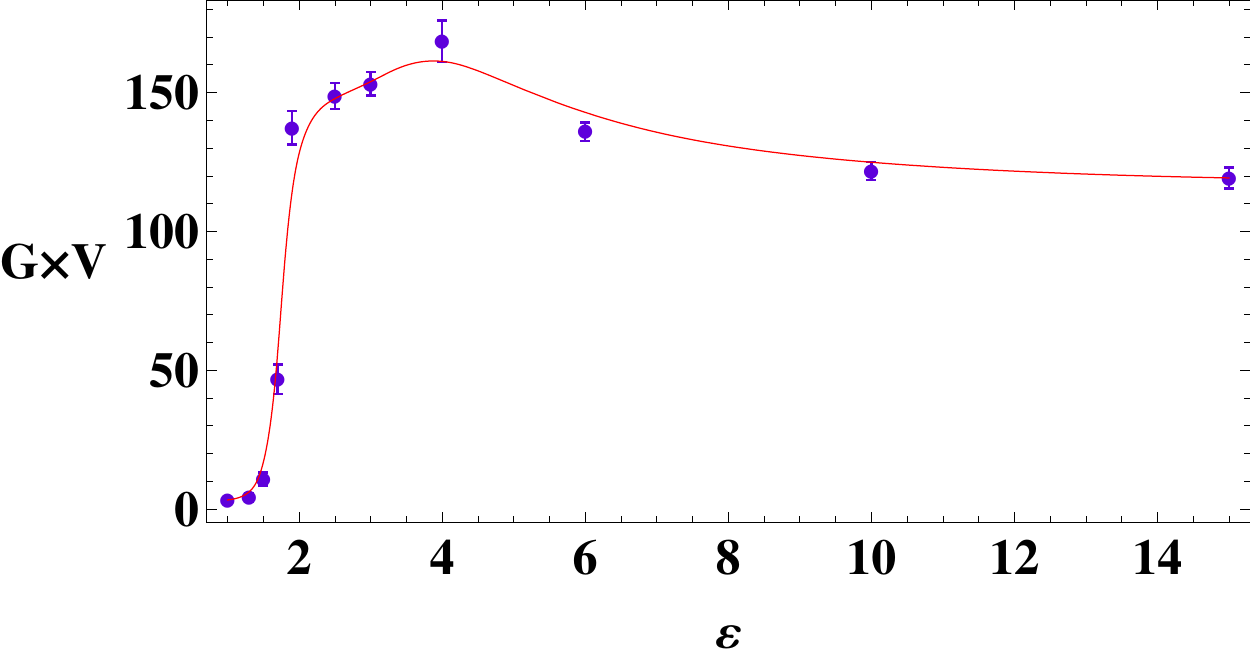}
  \caption{Rubber modulus (according to Eq. \ref{eq:eqxlink}) vs $\varepsilon$ for B3L12. At large $\varepsilon$, the high degree of association of the cluster suppresses the potential for connecting to other clusters - large clusters spaced further apart. Thus, the modulus of the network drops off at higher $\ve$. The solid line is a guide to the eye.}    
   \label{fig:modulus1}
  \end{figure}

\begin{figure}[ht]\centering
  \includegraphics[width=.5\textwidth]{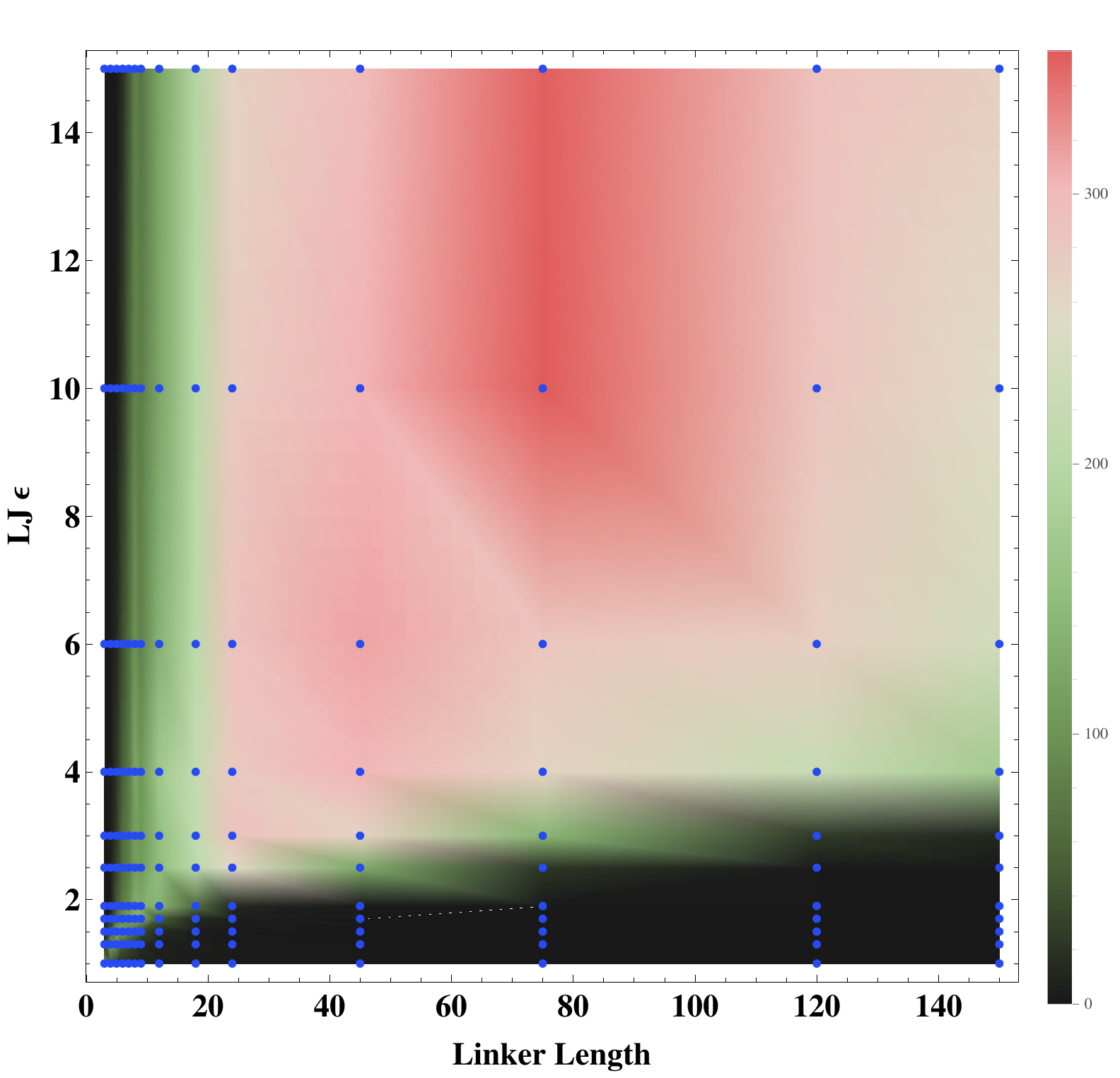}
  \caption{Overview of computed moduli for all molecular architectures studied here. Mechanical performance that is optimal in the sense that it provides the highest modulus arising due to a favorable balance between cluster size and cluster functionality is obtained for B3L75 and $\varepsilon=10$. }    
   \label{fig:densplot}
  \end{figure}

In the simplest approximation, we shall consider the effective network that emerges in the regime of intermediate $\ve$ and $N_l$ as a rubber with Guassian linkers connecting crosslinkers of varying functionality. In previous work, Yeo {\em et al.} \cite{yeo1981rubber} established that the shear modulus $G$ of such a network may be computed according to a modified classical rubber model (see, for instance, \cite{Rubinstein2007}) , as

\begin{equation}
\label{eq:eqflory}
G = g(f) \nu_e\, \kt
\end{equation}
In this equation,  $\nu_e$ is the number concentration of elastically effective chains, and g is the so-called front factor, which is to be determined according to the functionality distribution. In principle, $\nu_e$ counts both the contributions from physical entanglements ($\nu_p$) and crosslinkers ($\nu_c$), but since our network - outside of the network phase and at sufficiently low concentrations - does not possess entanglements that contribute to the modulus we will set $\nu_e=\nu_c$. The front factor $g(f)$ accounts for changes in density due to the contraction of the network upon crosslinking, as well as for the role of functionality. The combined result, obtained first by Yeo and based upon earlier work by Duiser and Staverman \cite{Duiser}, Graessly \cite{graessley1975statistical} and Tobolsky \cite{tobolsky1961rubber} then yields, for a network of average functionality $\bar f$

\begin{equation}
\label{eq:eqxlink}
G = \left(\frac{\bar f-2}{\bar f}\right)\left(\frac{\langle r^2 \rangle}{\langle r^2 \rangle_0}\right)\nu_c\, \kt
\end{equation}

in this equation $\frac{\langle r^2 \rangle}{\langle r^2 \rangle_0}$  is the ratio of the mean squared end-to-end distance of the polymer chains in the crosslinked network to that of the same chains in the uncrosslinked state. Becasue, as we have seen, primarily unit step length bridges occur we will count only those as elastically effective in the following. Using Eq. \ref{eq:eqxlink}, we may now estimate the shear modulus of the various network states in our simulations, using the functionality distribution to compute $\bar f$, and counting the number density of clusters to determine $\nu_c$. Sample results are presented in Fig. \ref{fig:modulus1}. The general trends we observe are consitent with what the phase diagram also suggests: As a function of both linker length and $\ve$, there exists and intermediate regime where cluster sizes are sufficiently large to permit high functionalities (many opportunities for connecting to other clusters), but are not yet so large that the number of potential partner clusters becomes limiting and the clusters become single collapsed entities. For shorter linker lengths (Fig. \ref{fig:modulus1}) the dropoff in modulus at higher values of $\ve$ is very pronounced, as the clusters quickly become fully isolated - for larger linker lengths (Fig. \ref{fig:modulus1}), the wider reach of every cluster (even when it is already fairly large) allows the system to retain some connectivity even at larger cluster sizes. Similar figures may be drawn for the dependence on the linker length, and in Fig. \ref{fig:densplot} we collect these into a modulus diagram. 

This diagram illustrates what we feel is a crucial point about these networks: more is not always better for increased rigidity. An optimum in modulus exists at intermediate $\ve$ as well as $N_l$, where the balance between connectivity opportunities and functionality is optimal for overall mecahnical response. It would be interesting to explore whether indeed such an optimal regime exists.
 
Finally we note that, obviously, fancier models are available to predict the mechanical properties of networks such as those we consider. Once fully formed, the structure of the hydrogel is no different from that of a telechelic gel, which was shown to be well-described by the classical theories of Flory \cite{paul1953principles} and Stockmayer \cite{stockmayer}, and whose kinetics of aggregation \cite{Baljon} and de-aggregation may be used to assess even their visco-elastic response \cite{Gucht1, Gucht2}.

\section{Computational rheology}
\label{sec7}

In order to verify the accuracy of rubber elastic models, we now turn to a direct, quantitative characterization of the viscoelastic response of the self-assembled network. Experimentally, establishing these properties is often challenging. There are various rheological methods, including micro- and macroscopic probing methods that can characterize the mechanical behavior of polymer networks.
 The macroscopic method is called bulk rheology. By this method one can relate the shear deformation (i.e., the strain $\gamma$), to the shear response (the stress $\sigma$) in a sample. We copy this protocol and implement it directly in our simulations. Fig. [\ref{fig:shear_snapshot}] shows a simulation box under oscillatory shear. For purely elastic response this relation is $\sigma=G \gamma$, where $G$ is called the shear modulus. For viscoelastic materials, that exhibit viscous responses as well as elastic responses, $G$ is usually decomposed into a real and an imaginary part:

\begin{figure}[ht]\centering
  \includegraphics[width=.5\textwidth]{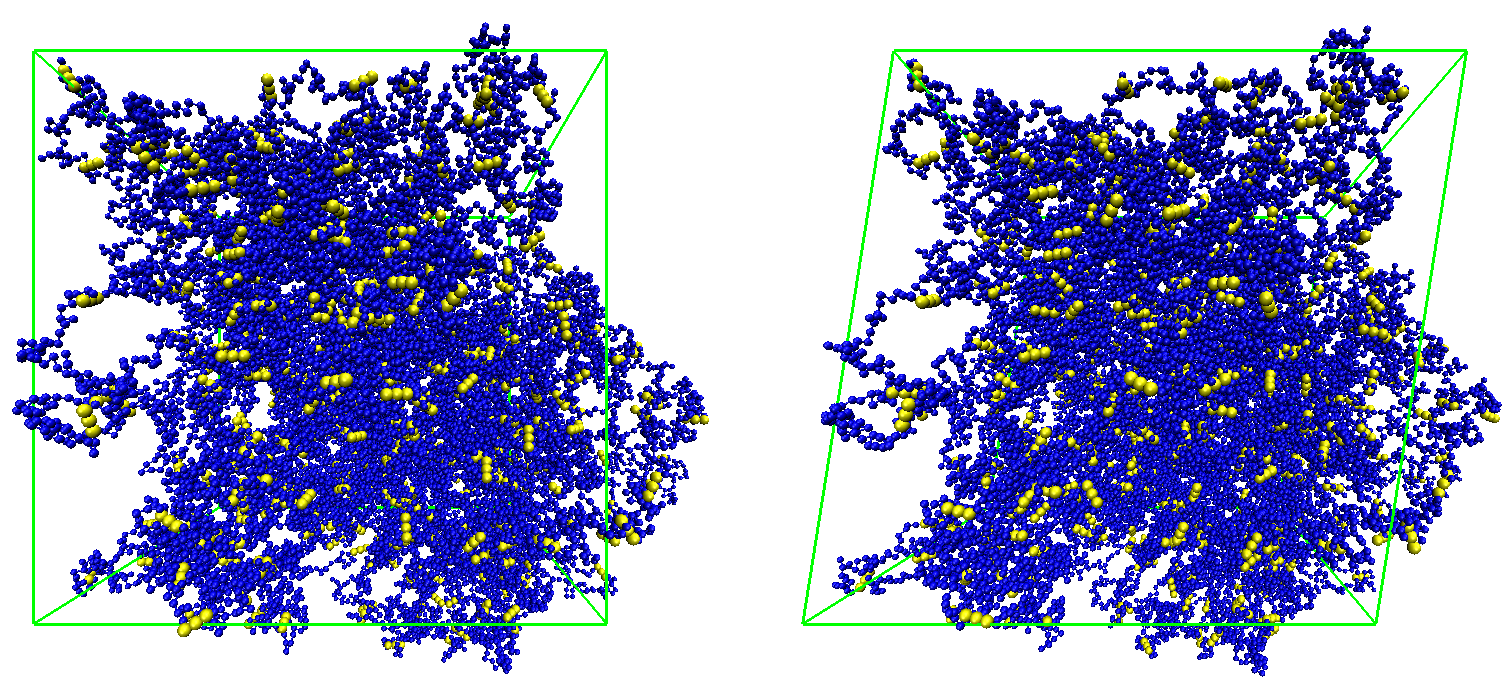}
  \caption{In  our oscillatory shear simulations the box is deformed (right) in the $xy$-plane, and executes a periodic oscillation characterized by a frequency $\omega$ and an amplitude $\gamma_0$. The network contained inside will accomodate this dynamic deformation differently depending on its ability - or inability - to undergo structural relaxations on the timescale of the applied strain.}\label{fig:shear_snapshot}
\end{figure}

\begin{equation}
G(\omega)\equiv G'(\omega)+iG''(\omega).
\end{equation}

\begin{figure}[ht]\centering
  \includegraphics[width=.5\textwidth]{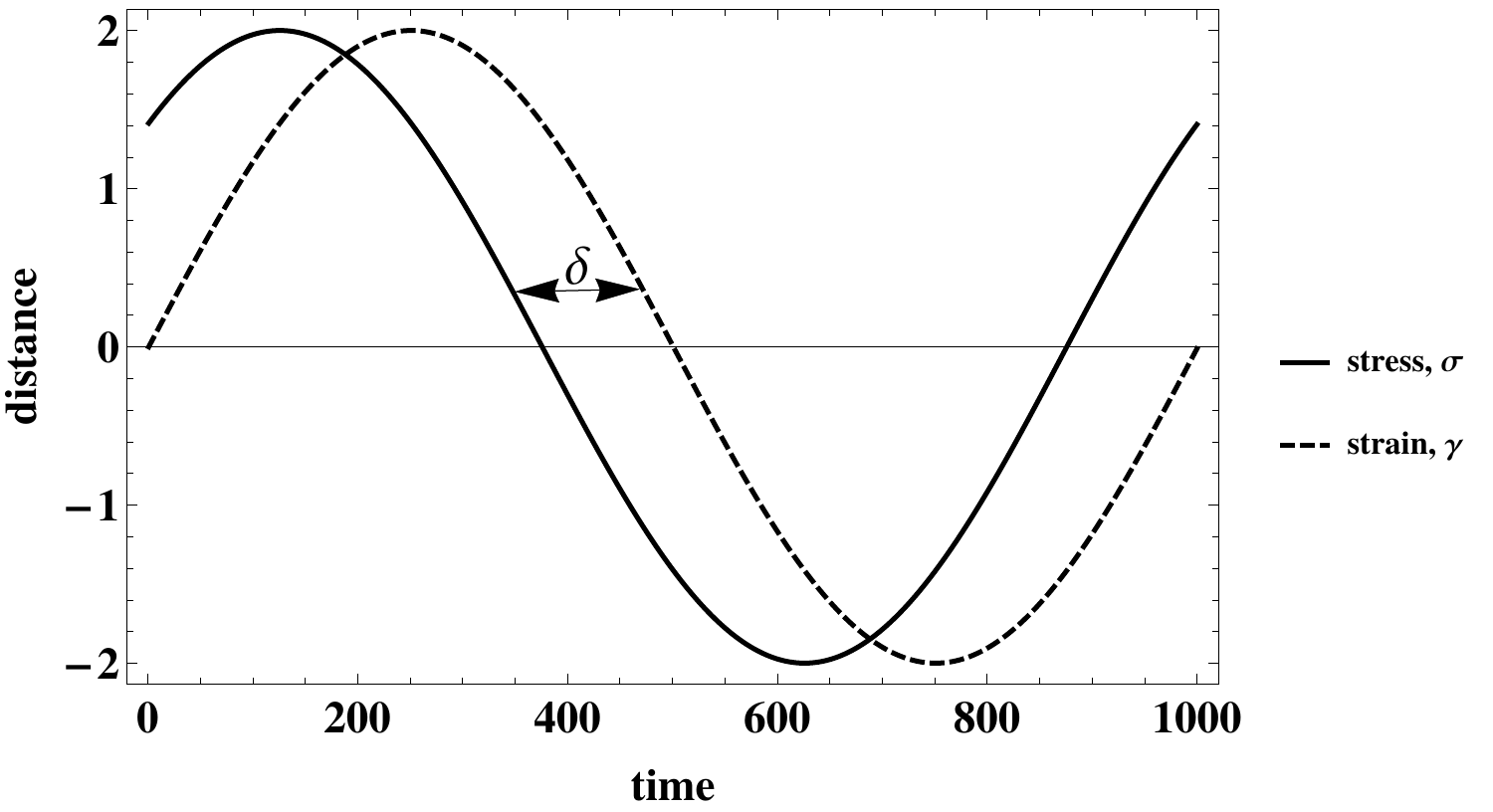}
  \caption{The phase lag between sinusoidal signal of deformation and response determines the elastic and viscous modulus. In this figure an example of $\delta=\frac{\pi}{4}$ is shown. This is an example of an ideal viscoelastic response $G'=  G^{\prime\prime} $.}\label{fig:ideal_visoelastic}
\end{figure}

In this equation, $G'$ is the {\em storage} modulus, that characterizes the elastic response of the material. The {\em loss} modulus $G''(\omega)$ quantifies the viscous response. For a purely Hookean solid, $G'(\omega)$ is constant, and $G''(\omega)=0$. For a Newtonian fluid, conversely, $G'(\omega)=0$ and $G''(\omega)=\omega \eta$ with $\eta$ the dynamic viscosity. In oscillatory rheology, a time-dependent strain of the form

\begin{equation}
\label{eq:oscillatory_shear}
\gamma=\gamma_0 \sin \omega t,
\end{equation}

where $\gamma_0$, amplitude, is the maximum deformation applied in each cycle. The time lag between two sinusoidal signals determines the viscoelastic moduli of the system:

\begin{align}
G'=\frac{\sigma}{\gamma}\cos \delta \quad \textrm{and} \quad G^{\prime\prime}=\frac{\sigma}{\gamma}\sin \delta.
\end{align}
In the above equations, $\delta$ is the phase lag between the stress and strain signals. The phase lag for a viscoelastic system is shown in Fig. [\ref{fig:ideal_visoelastic}].

In the case of linear response, this is the entire story - stress and strain always have the same frequency as no higher harmonics are generated. The only degree of freedom that linear response allows is the phase lag $\delta$, but in linear response the moduli themselves cannot in any way be functions of time, or of the strain amplitude, themselves. Since we will be interested in the non-linear (finite strain) response as well, it is useful to expand the stress response in the higher harmonics, expressing it as a Fourier series

\begin{align}
\label{eq:stress_func}
\sigma(t,\omega,\gamma_0)=\gamma_0 \sum_n &\lbrace G'_n(t,\omega,\gamma_0) \sin (n \omega t)\\ + &G^{\prime\prime}_n(t, \omega,\gamma_{0}) \cos (n \omega t \rbrace\,.
\end{align}

Where $G'_n$ is storage modulus, a measure of elastic energy stored in the material. Provided the applied strain is sinusoidal, $G'_n$ indicates the magnitude of the $n^{\rm th}$ harmonic of the stress response. The loss modulus $G^{\prime\prime}_n$ is correlated with the viscous dissipated response and measures the out-of-phase component of the stress. If one has the full signals $\sigma(t)$ and $\gamma(t)$, all the moduli $G_n$ can be determined from the complex coefficients $c_n$ of the discrete Fourier transform of the discrete stress time series $\sigma_t$

\begin{equation}
c_n= \frac{1}{N}\sum_{t=0}^{N-1}\sigma_t e^{-i2\pi nt/N},
\end{equation}

where $N$ is the period of the applied strain times the sampling frequency. Using the relationships between the Fourier coefficients $a_n=c_n+c_{-n}$ and $b_n=i(c_n+c_{-n})$, along with knowledge that $c_{-n}$ is the complex conjugate of $c_n$, each harmonic modulus can then be determined as 

\begin{equation}
\label{eq:harmonicG}
G'_{n}=\frac{b_n}{\gamma_0}=\frac{2\Im(c_n)}{\gamma_0}\quad \textrm{and}\quad G^{\prime\prime}_n=\frac{a_n}{\gamma_0}=\frac{2\Re(c_n)}{\gamma_0}.
\end{equation}

With $\Im(c_n)$ and $\Re(c_n)$ being the imaginary and real parts of the complex Fourier coefficients respectively. To assess whether or not a particular system, exposed to a strain with period $\omega$ and amplitude $\gamma_0$ can be considered in a regime of linear response, we typically monitor the ratio of the first to the third harmonic modulus (the second is zero, by symmetry).
\subsection{Model and simulation protocol}
\label{subsec:subs1}

 As stated in section \ref{sec2}, in all of  simulations so far we have studied a single, very long chain consisting of 500 repeats of the binder+linker motif. In the experiments, chains are typically much shorter. We expect there to be little difference between studying one long chain, or chopping this long chain up into many smaller chains. Now, we  verify this explicitly: we compare the results of one repeating chain of $500$ binder-linker groups with those obtained for $27$ individual chains of $19$ binder-linker groups each (this is the typical repeat number for the experimentally used Poly(8kU4U) molecule). Both systems are allowed to self-assemble into a crosslinked network.  Each binder is made up of three spherical beads and every linker is made up of $N$ beads, and we  let $N$ vary corresponding to the length of the linker from $3$ to $150$ beads. To  ensure a fixed bond length in  our simulations, we  choose a large value for the strength of the bond potential $k_b=400 \, \kt / \sigma^2$.  The hydrophobic chains in the experimental system are quite short, roughly $5 nm$, which makes them act like a stiff rod \citep{pawar2012injectable}. To simulate this we  choose a large value for the strength of bending potential $k_a=50\, \kt$. We  also compare the results of simulations for harmonic bond potential with FENE bond potential \citep{bird1977dynamics} by which the nearest-neighbour beads along the chain interact through an anharmonic, finitely extensible, non-linear, elastic potential given by

\begin{equation}
\label{eq:fene}
U_{FENE}(r)=-\frac{1}{2}k R_0^2 \ln \Bigg[ 1- \bigg( \frac{r}{R_0}\bigg)^2 \Bigg], \quad \textrm{for} \quad r<R_0.
\end{equation}

The parameter values of $R_0=1.2$ and $k=100$ prevent entanglement and overlapping of chains \citep{baschnagel2005computer}. The FENE potential diverges logarithmically as $r \rightarrow R_0$, providing a finite distance between chain beads. It is used in coarse grained simulations to encode the finite extensibility of real polymers.

According to the above equation, particles closer than $\sigma$ interact via a repulsive Lennard-Jones potential whereas beyond $\sigma$ the interaction is zero to ensure real chain scaling - See also section \ref{sec2}. We  study the effect of temperature and density on the mechanical properties of the system and compare them with the experimental results. To simulate experiments at different concentrations, we  keep the number of the beads in simulation constant, but change the size of the simulation box to decrease or increase the concentration. To do so, we  perform NPT simulations: here, we  dial in the pressure to control the box volume. We note, that this may induce other effects besides those of concentration alone, as we  necessarily have to go to elevated pressures to realize higher concentrations. We  have not studied these prestress effects separately. We choose three different initial pressures to work with: $P=$ $200$, $250$ and $300$ (molecular dynamics units\footnote{The molecular dynamic units and their conversion to real units are discussed in Appendix A}). Our procedure for the equilibration and determining the pressure and volume of the system is as follows:

\begin{figure}[ht]\centering
  \includegraphics[width=.5\textwidth]{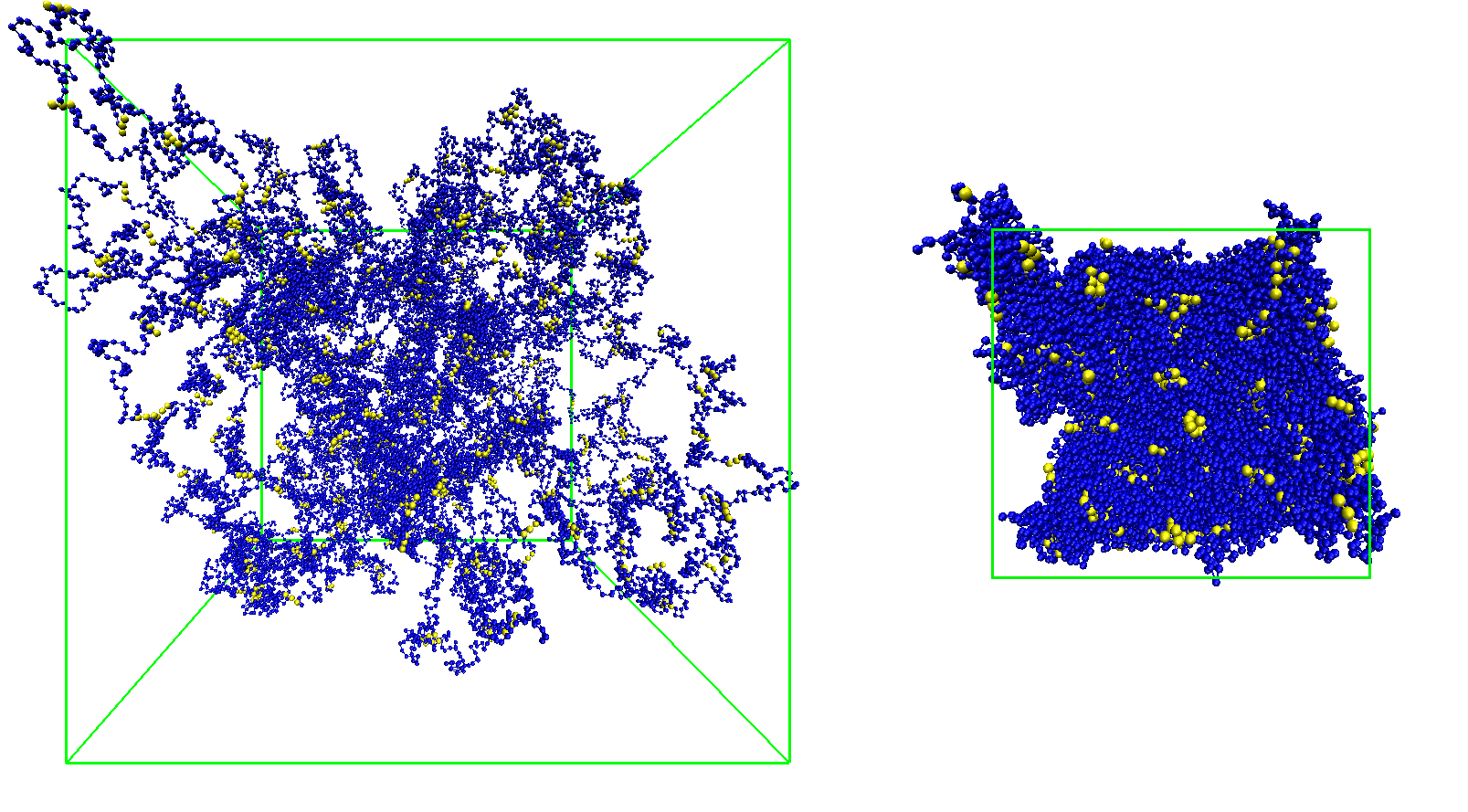}
  \caption{Left: the system as it is started off, from a random initial geometery in a big box. Right: the pressurized system, self-assembled in an NPT run}\label{fig:NPT}
\end{figure}

 All simulations start from a random geometry in an enormously large box. In the first stage of the simulation, we  perform a NVT simulation, that is coupled to a Nos\'e -Hoover thermostat, in a large box to thermostat the temperature of the system to a desired temperature (figure \ref{fig:NPT}). In the next stage, we  shrink the simulation box to a very high density via a NPT ensemble simulation while also self-assembly of the block copolymers takes place. Once the temperature and the total energy of the system become stationary, we  perform a series of NPT runs to expand the box until the pressure is just zero. This is the volume at which the chains precisely fit inside the box, but do not exert any outward (or inward) pressures on it. We take this volume to be the proper volume of a particular self-assembled configuration, and use it to determine the density. This density, $\phi^{\ast}$, corresponds - though likely not identical - to the overlap concentration in experimental systems which is:

\begin{equation}
C^{\ast}\sim\frac{N}{R_g^3}\,,
\end{equation} 

where $N$ is degree of polymerization and $R_g$ is radius of gyration of polymer \citep{daoud1975solutions}. Since the number of beads in  our simulations do not vary, we  can say that $C/C^{\ast}=\phi / \phi^{\ast}$. This run is followed by more NPT runs which  the simulation box is compressed to provide higher density systems.
 
The rheology simulations are performed with periodic boundary conditions using the NVT ensemble, which serves to ensure that the pressure of the system changes corresponding to the applied deformation - in LAMMPS the stress must be determined from the pressure tensor. We solve the SLLOD equations of motion which were proven to be equivalent to Newton's equations of motion for shear \citep{evans1984nonlinear}.  In this method, instead of boundary driven deformation , where shearing deformation is induced to the particles by the motion of the boundaries, a velocity gradient is generated to move all the particles proportional to the deformation of the boundaries. These equations are coupled to a Nos\'e -Hoover chain thermostat in a velocity Verlet formulation. The oscillatory shear strain - according to Eq. (\ref{eq:oscillatory_shear}) - is applied to the simulation box in the $xy$-plane, for various amplitudes and oscillation periods. To obtain the shear stress response, we  compute the $xy$ component of the symmetric (Cauchy) stress tensor, $\sigma_{xy}$, for every bead in the simulation box and sum over all atoms every $5$ simulation time-steps. In the following sections we  investigate the feasibility of performing rheology with the LAMMPS molecular dynamics package \citep{plimpton1995fast}.

\subsection{Computing stress and temperature control}
\label{subsec:subs2}

The oscillatory shear simulation imparts a continuously changing deformation to the simulation box. As a result, for affine deformation, each atom (bead) in the simulation box can be thought of as being forced to drift at a given velocity. For example, if the box is being sheared in the $xy$-plane the atoms at the bottom of the box (low $Y$) have a smaller velocity in the $x$-direction than those atoms at top of the box (at high $y$). LAMMPS subtracts this spurious position-dependent drifting velocity from each atom while shearing.

To obtain the viscoelastic modulus of a molecular system using Eq. (\ref{eq:stress_func}), one needs to compute the stress response of each particle inside the system to external deformation. The Cauchy stress tensor, a $3$ by $3$ tensor, completely defines the state of stress at any particular point of a material structure in a given deformed state. Written out in components, the Cauchy stress tensor has the following form:

\begin{equation}
{\sigma}=
\left[{\begin{matrix}
\sigma _{xx} & \sigma _{xy} & \sigma _{xz} \\
\sigma _{yx} & \sigma _{yy} & \sigma _{yz} \\
\sigma _{zx} & \sigma _{zy} & \sigma _{zz} \\
\end{matrix}}\right]
\, .
\end{equation}

\begin{figure}[ht]
\centering
	\subfigure[$\gamma=5\%$]	
	{\includegraphics[width=0.20\textwidth]{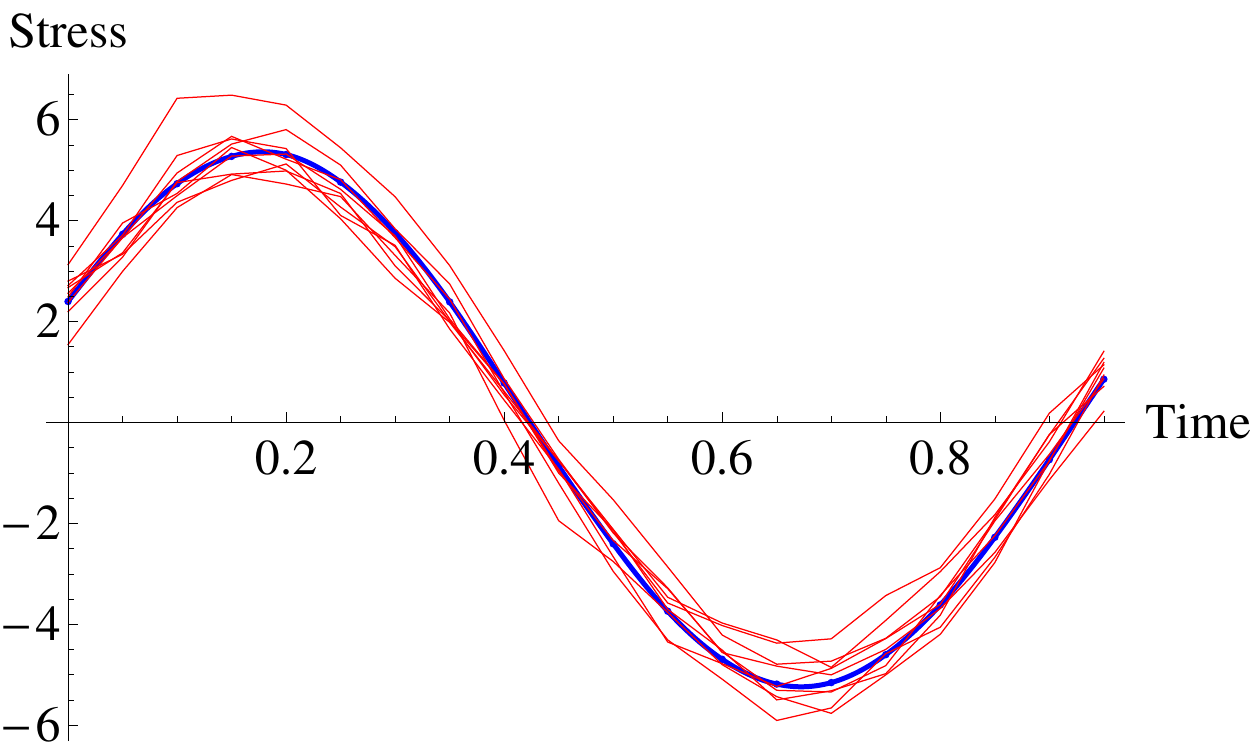}}
	\subfigure[$\gamma=10\%$]
	{\includegraphics[width=0.20\textwidth]{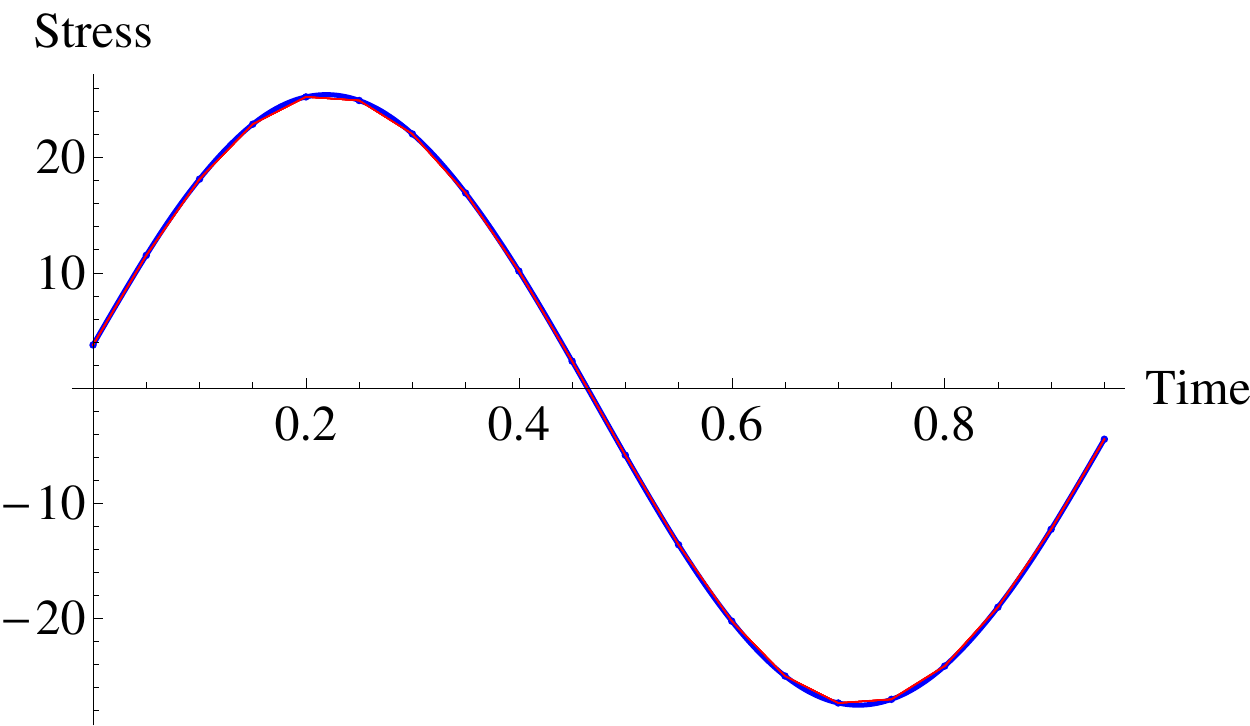}}
	
\caption{Stress fluctuations. Here we show the stress response over one period of the strain, for two different maximal shear values. The stress response is drawn in red every $50$ cycles, and the blue line is the average of all the cycles. (a) The shear rate is slow compared to fluctuation timescales, so the stress response is slightly different in every cycle. (b) For higher deformation, larger than $\gamma \approx 8\%$, the system becomes more rigid and fluctuations are suppressed.}
\label{fig:shear rate}
\end{figure}

Thus, $\sigma_{ij}$ is the force in the $i$-direction on the surface whose normal is in the $j$-direction. The diagonal components of the stress tensor are force per area in all three dimensions, and contain the hydrostatic pressure as well as any normal stresses that develop inside the material. In  our simulations, the complete stress tensor for atom $i$, multiplied by the volume that the atom occupies, in  our simulations is computed as follows:

\begin{align}
\label{eq:stress_tesor}
\sigma_{ij}=-[mv_iv_j &+ \frac{1}{2}\sum_{n=1}^{N_p}(r_{1i}F_{1j} + r_{2i}F_{2j})\nonumber \\
 &+ \frac{1}{2}\sum_{n=1}^{N_b}(r_{1i}F_{1j} + r_{2i}F_{2j})\nonumber \\
  &+ \frac{1}{3}\sum_{n=1}^{N_a}(r_{1i}F_{1j} + r_{2i}F_{2j}+r_{3i}F_{3j})].
\end{align}

\begin{figure}[ht]\centering
  \includegraphics[width=.5\textwidth]{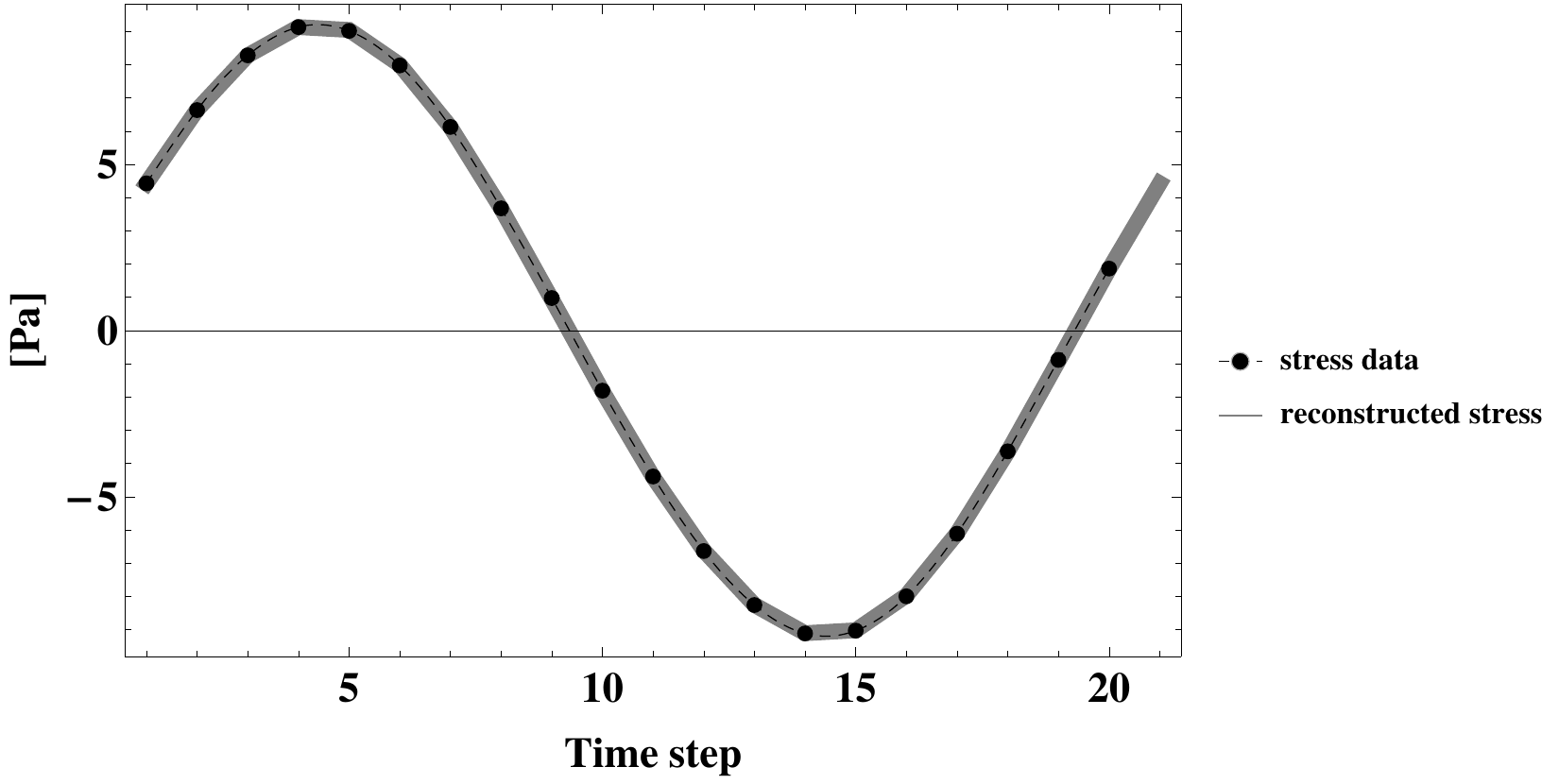}
  \caption{The stress signal for $\gamma=8\%$, $t=4$ and $\omega=0.2$, reconstructed from only the first harmonic of the full Fourier expansion fits very well to the numerical stress data, evidencing that we am in a regime of linear response.}\label{fig:recons}
\end{figure}

The first term is a kinetic energy contribution for atom $i$. The second term is a pairwise energy contribution where $n$ runs over the $N_p$ neighbors of atom $i$ which are all the atoms that are within the cut-off range of the potential, $r_1$ and $r_2$ are the position of the two atoms involved in the pairwise interaction, and $F_1$ and $F_2$ are the forces on the $2$ atoms resulting from the pairwise interaction. The third term is a bond contribution of similar form for the $N_b$ bonds which atom $i$ is part of. There is also a similar term for the $N_a$ angle interactions that atom $i$ is involved in. The so-called virial stress tensor is similar to the Cauchy stress tensor, and has all the above terms except for the contribution from the kinetic energy. In  our systems we find no significant difference in the moduli calculated by these two different methods of computing the stress tensors, because the actual velocities remain low and the energy is dominated by the bonded and non-bonded contributions. The comparison is presented in Fig. [\ref{fig:vsvirial}]. As discussed in section \ref{subsec:subs1}, we convert the stress signal to dynamic moduli using discrete Fourier function. In the linear regime only the first Fourier harmonic of the stress signal has contribution to the modulus. To check that the calculated modulus is in the linear regime we reconstruct the stress signal from only the first Fourier harmonic using the following equation

\begin{equation}
\sigma(t)= G^{\prime \prime}_{0} \frac{\gamma_{0}}{2}+ \gamma_{0} \sum_{n=1,3,5,\dots} \left( G'_n \sin (2\pi nt) +  G^{\prime\prime}_n \cos(2\pi nt)\right),
\end{equation}

where $ G^{\prime\prime}_{0}$ is the $n=1$ component of Eq. (\ref{eq:harmonicG}). The reconstructed stress, obtained from the above formula, fits very well to the stress data from simulations. This is shown in the Fig. [\ref{fig:recons}].

\begin{figure}[ht]\centering
  \includegraphics[width=.45\textwidth]{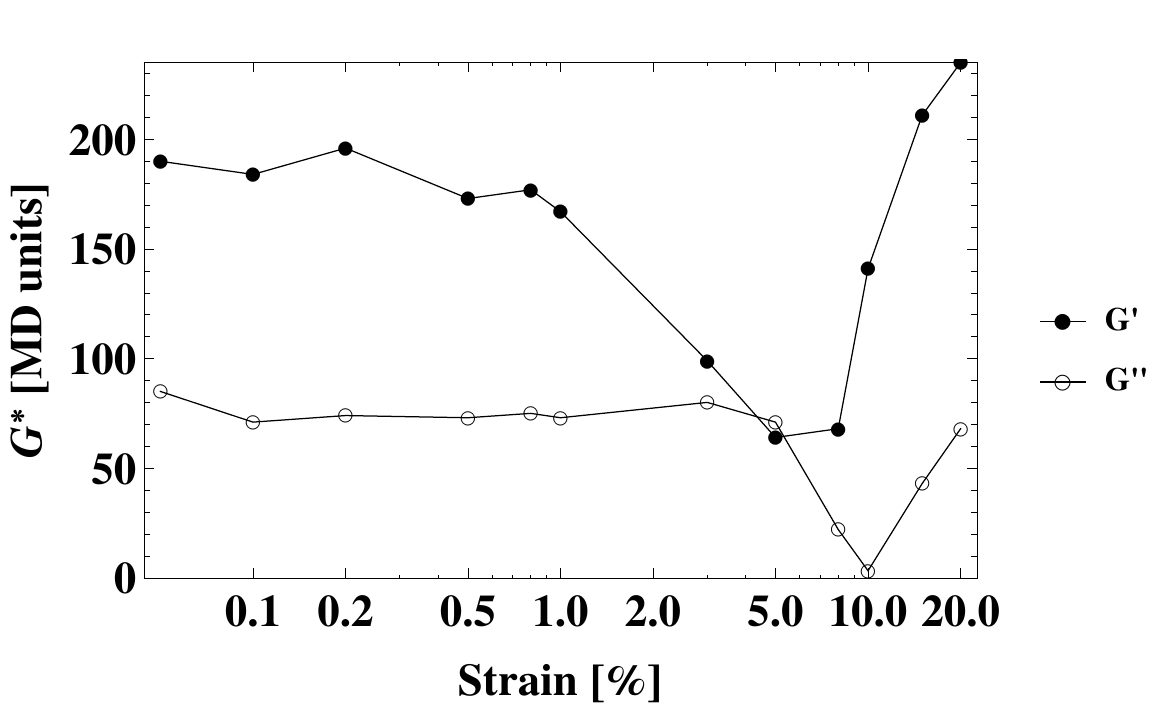}
  \caption{Strain dependent modulus for B3L45, $t=4$, $\phi/\phi^{\ast}=2.7$. The rise of modulus in the high shear rate region is due to bond stretching and hard core potential of the beads during the simulation. This regime is unphysical - at such high strain rates the implicit solvent interactions break down and  our simulations can no longer capture the actual deformations of the material. We discard points after the minumum, and interpret the rapid dropoff as yielding behavior for  our system.}\label{fig:yield}
\end{figure}

In molecular dynamics simulation of oscillatory shear, care should be taken in choosing the right deformation rate. If the box deformation rate is larger than the time-scale in which the beads interact and fluctuate then the position and interaction of the beads are the same in all of the oscillatory cycles (see Fig. [\ref{fig:shear rate}]) and the response is similar to the response of a crystalline materials which has significantly higher modulus than soft materials. Obviously increasing the shear rate eventually would lead in stretching the bonds more than equilibrium bond distance which breaks the bonds and stops the simulation. In the Fig. [\ref{fig:yield}] is shown that the mechanical response of the system increases for $\gamma>8$ which is towards suppressing the fluctuations and solid-like structures. This rise in modulus eventually terminates at $\gamma=25$ where the simulation stops because of excessive bond stretching.

In MD simulations a thermostat must be used as a means of controlling the temperature of particles. Typically a target temperature $T$ is specified by the user, and the thermostat attempts to equilibrate the system to the requested $T$. To compute the temperature, the kinetic energy is divided by the Boltzmann constant and $n_{\rm dof}$, number of degrees of freedom present in the system:
\begin{align}
\label{eq:temperature}
E_{\rm kin}&= \frac{1}{2} \kt n_{\rm dof} \nonumber \\
T & \equiv \frac{2 E_{\rm kin}}{\kt n_{\rm dof} }.
\end{align}
Since the kinetic energy is a function of particle velocity, there is often a need to distinguish between a particle's {\em streaming velocity} which occurs due to  group motion of aggregated particles and its {\em thermal velocity} due to thermal fluctuation. The sum of the two is the particle's total velocity. Using the Nos\'e -Hoover equations \citep{nose1984unified} we thermostat the translational velocity of particles and subtract a velocity bias that is the result of deforming the simulation box. Hence the dependence of the computed stress on temperature is only the contribution of thermal velocity which is due to the fluctuations of the particles.

\begin{figure}[ht]
\centering
	\subfigure[]	
	{\includegraphics[width=0.23\textwidth]{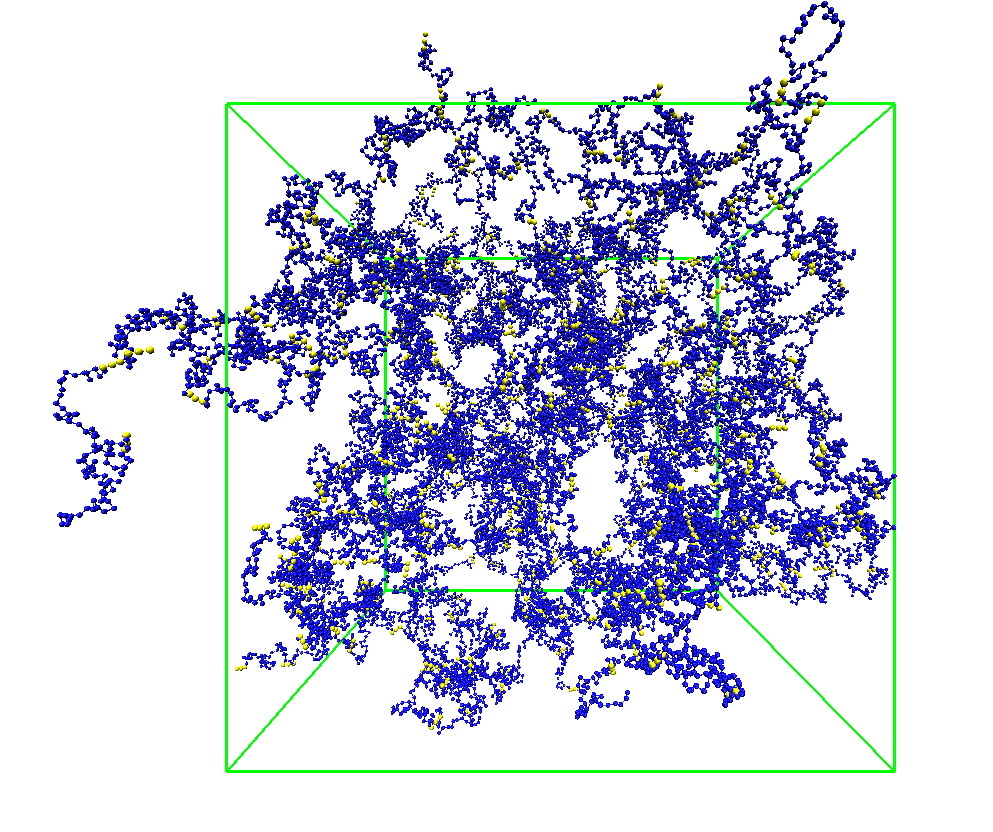}}
	\subfigure[]
	{\includegraphics[width=0.23\textwidth]{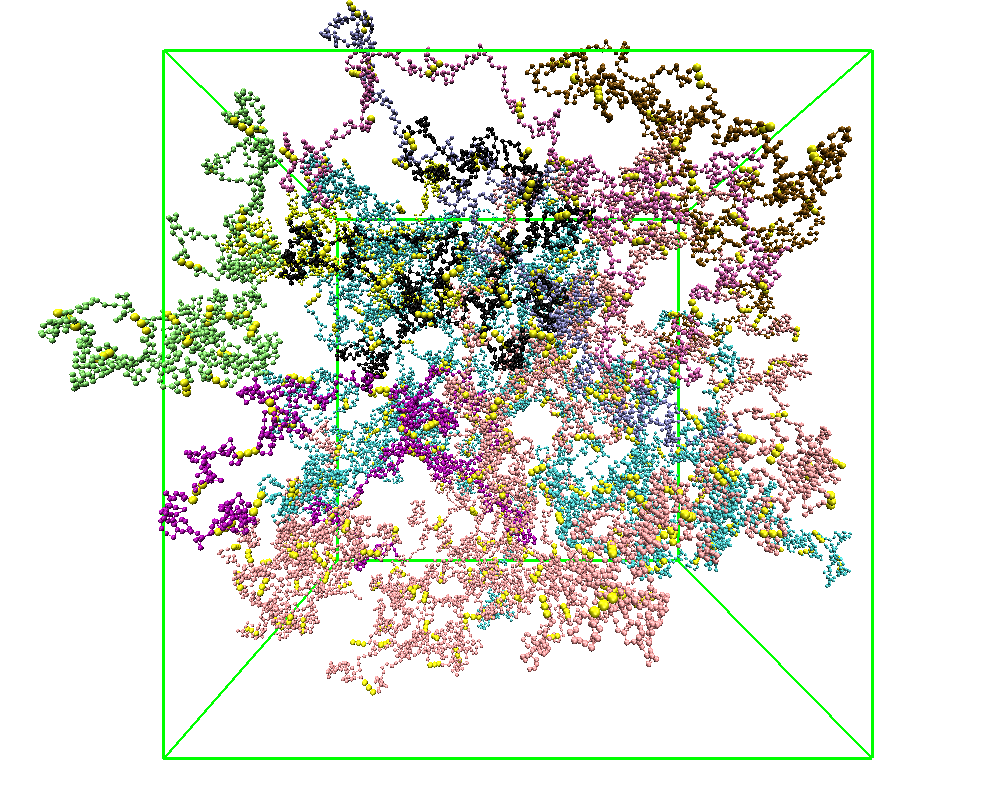}}
	\subfigure[]
	{\includegraphics[width=0.45\textwidth]{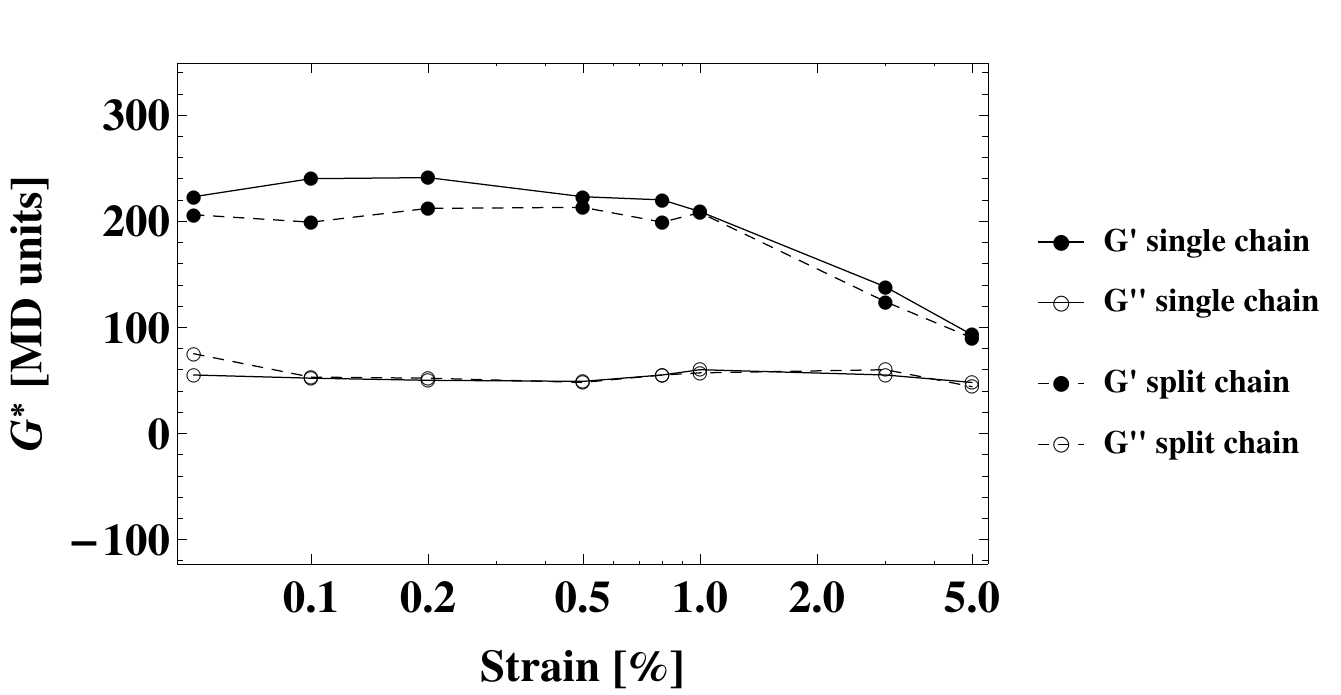}}
	
\caption{Simulation snapshots of (a) single long chain with 500 repeating binder+linker blocks and (b) split chain, where the separate chains are distinguished by different colors.(c) Strain-dependent simulation results for two different systems of  single chain (solid line) and split chain (dashed line). Storage modulus is shown in blue color and loss modulus is shown in red color. As may be seen here, the results are identical for the split and the long system.}
\label{fig:vssplit}
\end{figure}

\section{Results}\label{sec:rheology/results}

Now, we present the results of a series of oscillatory rheology simulations of the self-assembling multiblock copolymer system. First, we study the effect of changing the molecular architecture from a single long chain to several short chains: Strain-controlled simulations results at different densities are shown in Fig. [\ref{fig:vssplit}]. To justify the validity of  our coarse-grained model to simulate a block copolymer, we compare the results of two different system, $(i)$ a one repeating chain that is self-assembled to a flowerlike network $(ii)$ split chains which are $27$ chains of $19$ hydrophobic-hydrophilic block each. The system sizes in both cases are equal to $24000$ particles. Fig. [\ref{fig:vssplit}] shows equivalent results for both systems.

\begin{figure}[ht]\centering
  \includegraphics[width=.5\textwidth]{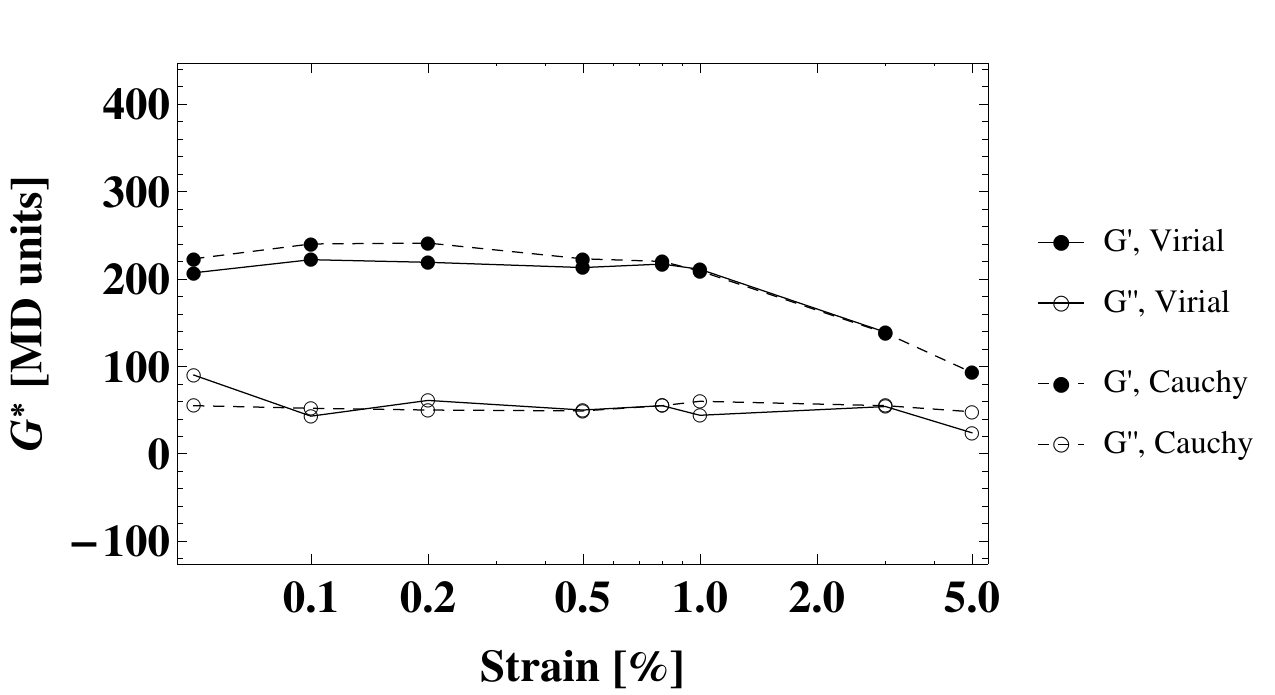}
  \caption{Strain-dependent modulus of a of B$3$L$45$ crosslinked network at $\phi/\phi^{\ast}=2.7$, $\omega=1$. The complex modulus was calculated from the two different stress tensors: Cauchy and Virial.}\label{fig:vsvirial}
\end{figure}

Second, we demonstrate the practical equivalence (in  our simulation settings) for the virial and Cauchy stress approaches \citep{zhou2003new,zimmerman2010material}. Here we compare modulus of one system using Cauchy and Virial stresses. As shown in Fig. [\ref{fig:vsvirial}], the difference of two stresses in  our model is small. 

\begin{figure}[ht]
\centering
	\subfigure[Harmonic spring bond, $\phi/\phi^{\ast}=2.7$]
	{\includegraphics[width=0.45\textwidth]{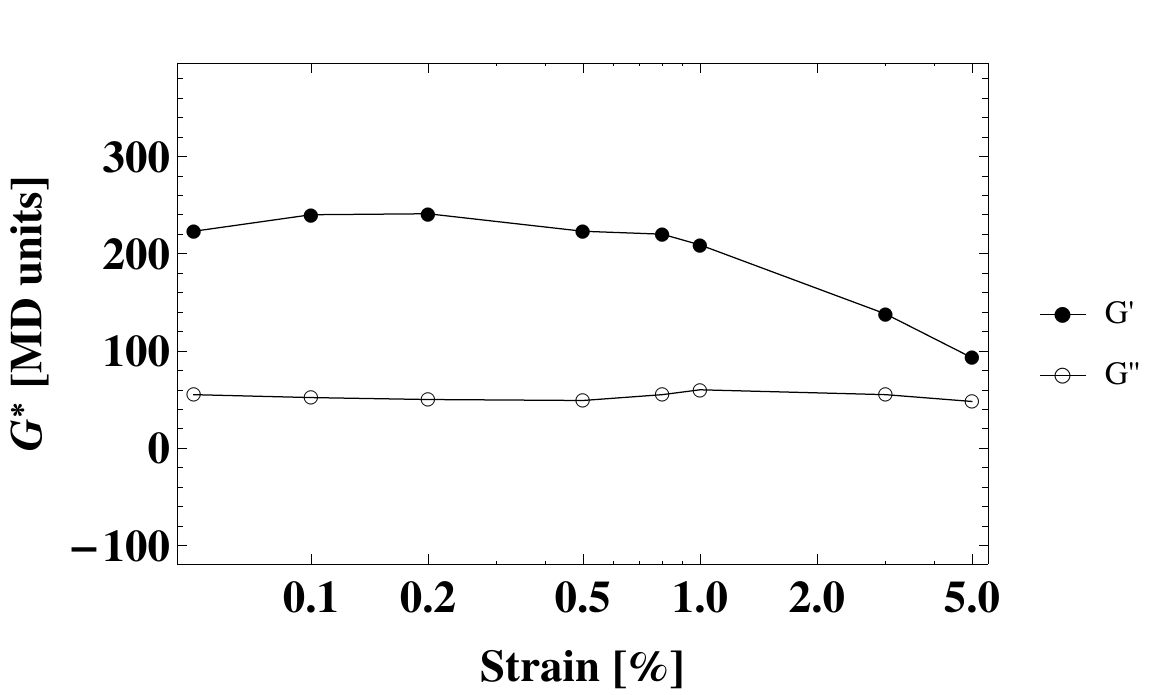}}
	\subfigure[FENE bond, $\phi/\phi^{\ast}=2.7$]
	{\includegraphics[width=0.45\textwidth]{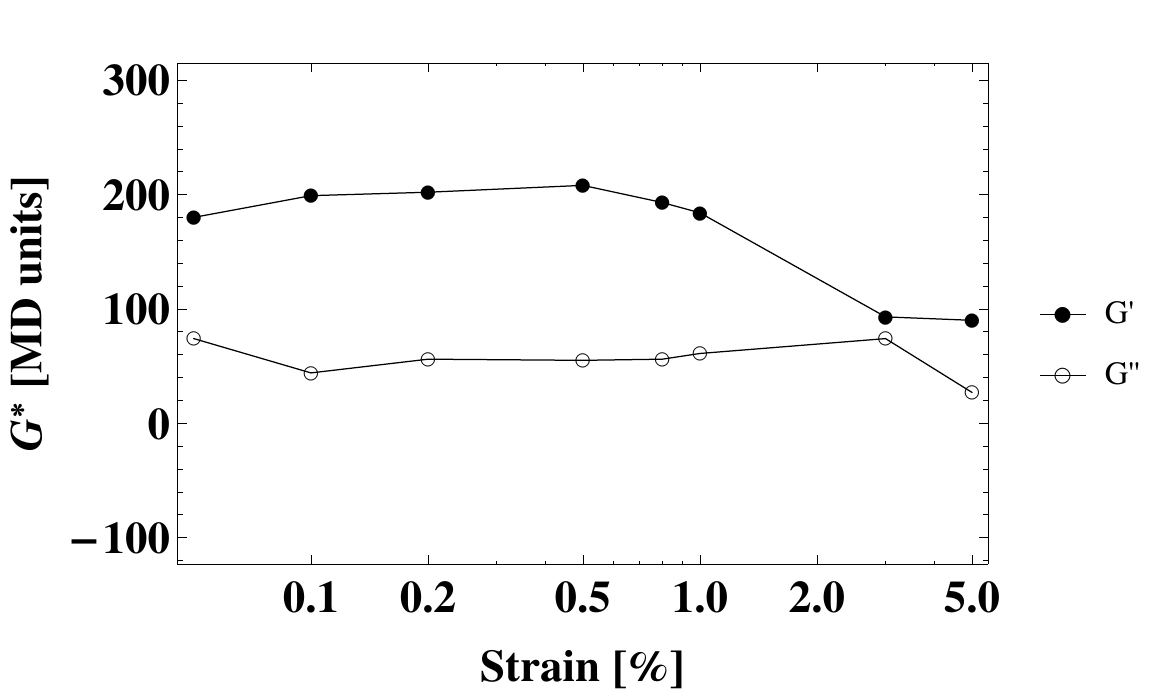}}

\caption{Strain-dependent comparison between a bead-spring polymer network of harmonic bonds vs a bead spring network of FENE bonds for B$3$L$45$, $\omega=1$ and $T=2$. The storage modulus is shown in solid symbols, the loss modulus is shown in open symbols.}
\label{fig:vsfene}
\end{figure}

Finally, we examine the effects of the details of the bond potential. As discussed in  section \ref{sec:rheology/results}, FENE bonds are often used as finitely extensible bonds in coarse grained MD simulations to better reflect the finite backbone length of polymers. To compare the effect of harmonic bonds vs. FENE bonds in the shear deformation simulations we compare the modulus for two systems containing two above bonds in Fig. [\ref{fig:vsfene}]. Again, the results are completely similar: This may be understood from the fact that in the regime where we measure, the non-linear stretching regime is never engaged and the FENE bonds act as linear springs.

\begin{figure*}[ht]
\centering
	\subfigure[, $\phi/\phi^{\ast}=2.7$]
	{\includegraphics[width=0.3\textwidth]{b_p200-eps-converted-to.pdf}}
	\subfigure[ $\phi/\phi^{\ast}=2.9$]
	{\includegraphics[width=0.3\textwidth]{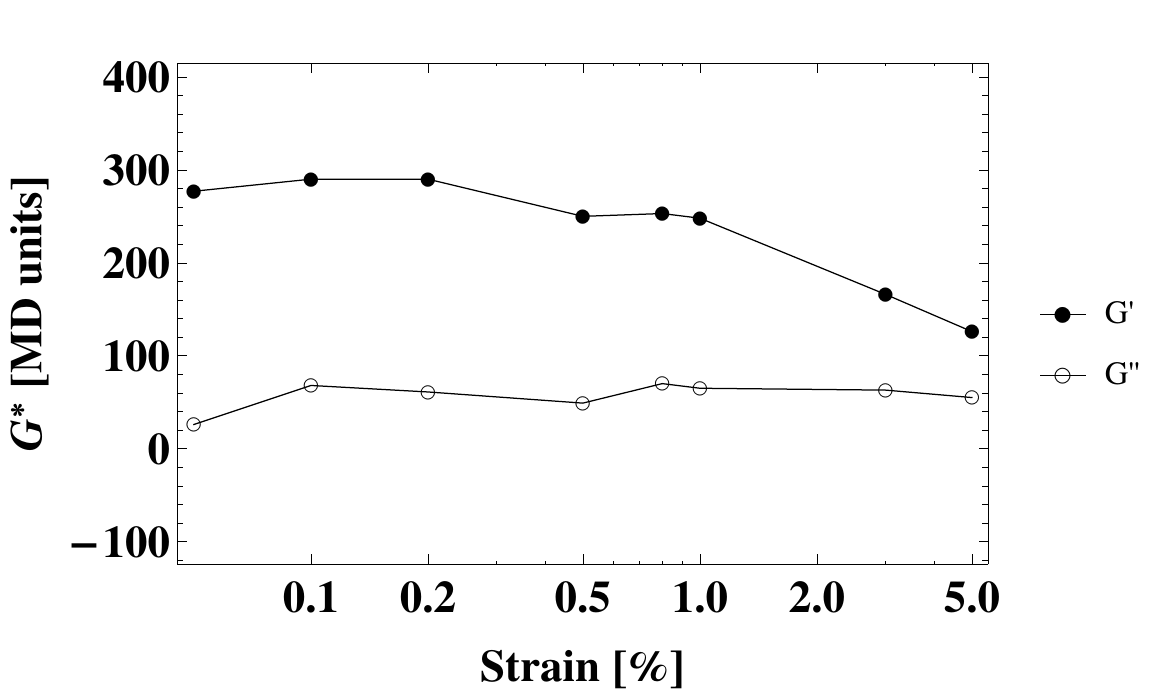}}
		\subfigure[ $\phi/\phi^{\ast}=3.1$]
	{\includegraphics[width=0.3\textwidth]{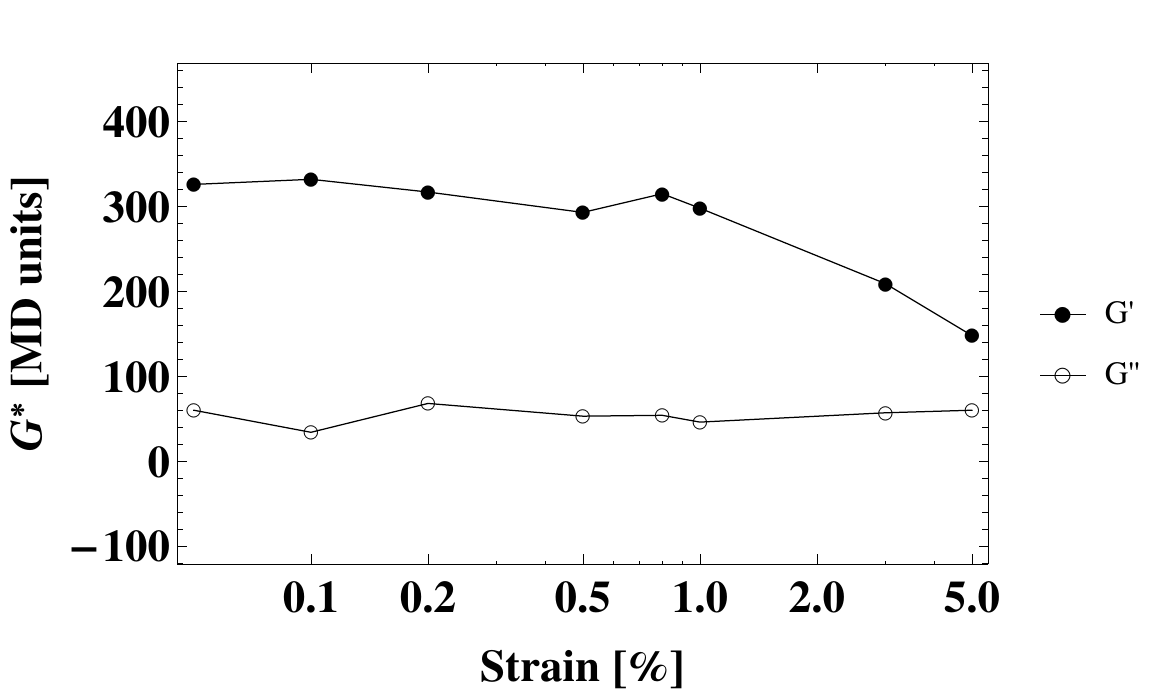}} 
	
\caption{Dynamic moduli $G'$ and $G''$, measured in strain-controlled settings at $\omega=0.2$ and $T=1$ at various concentrations. The storage modulus is shown in solid symbols, the loss modulus is shown in open symbols.}
\label{fig:vsconcentration}
\end{figure*}

The effect of concentration is also shown in Fig. [\ref{fig:vsconcentration}]. The overlap concentration - which is the concentration at which the single molecule network precisely fits within the box $\phi^{\ast}$ - is obtained by relaxing the network volume until the pressure first reaches zero. Higher concentrations are shown in units of this overlap concentration.

\newpage

\begin{figure*}[t]
\centering
	\subfigure[$T=1,\phi/\phi^{\ast}=2.7$]{\includegraphics[width=0.45\textwidth]{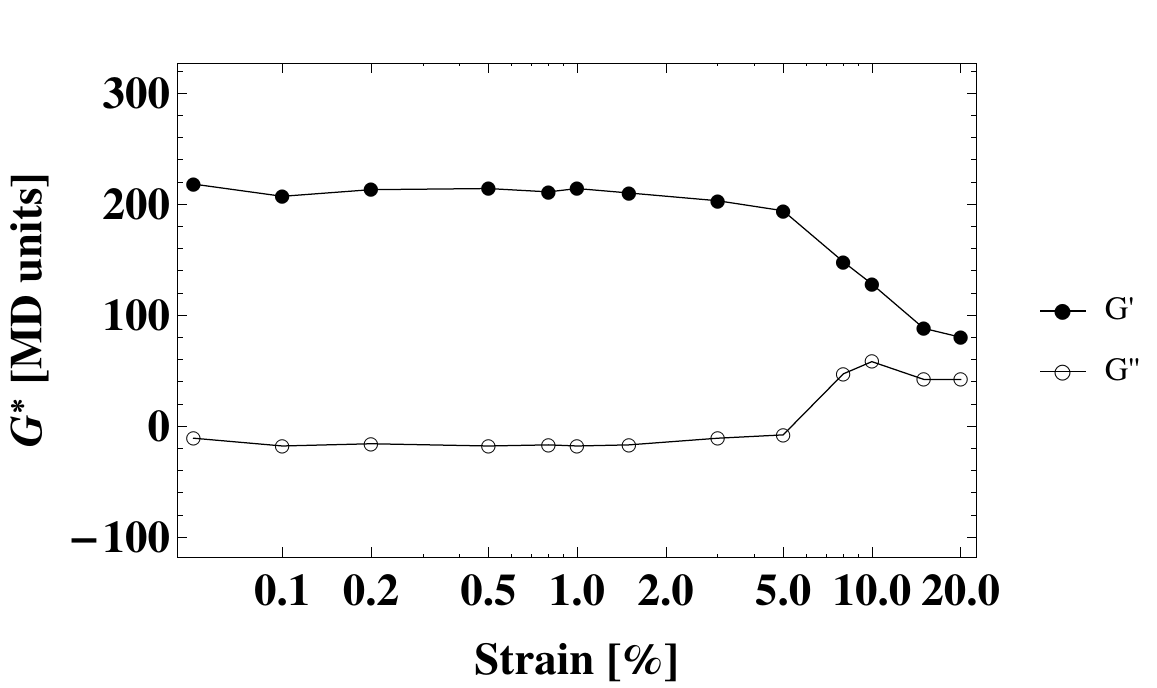}}
	\subfigure[$T=2, \phi/\phi^{\ast}=2.7$]{\includegraphics[width=0.45\textwidth]{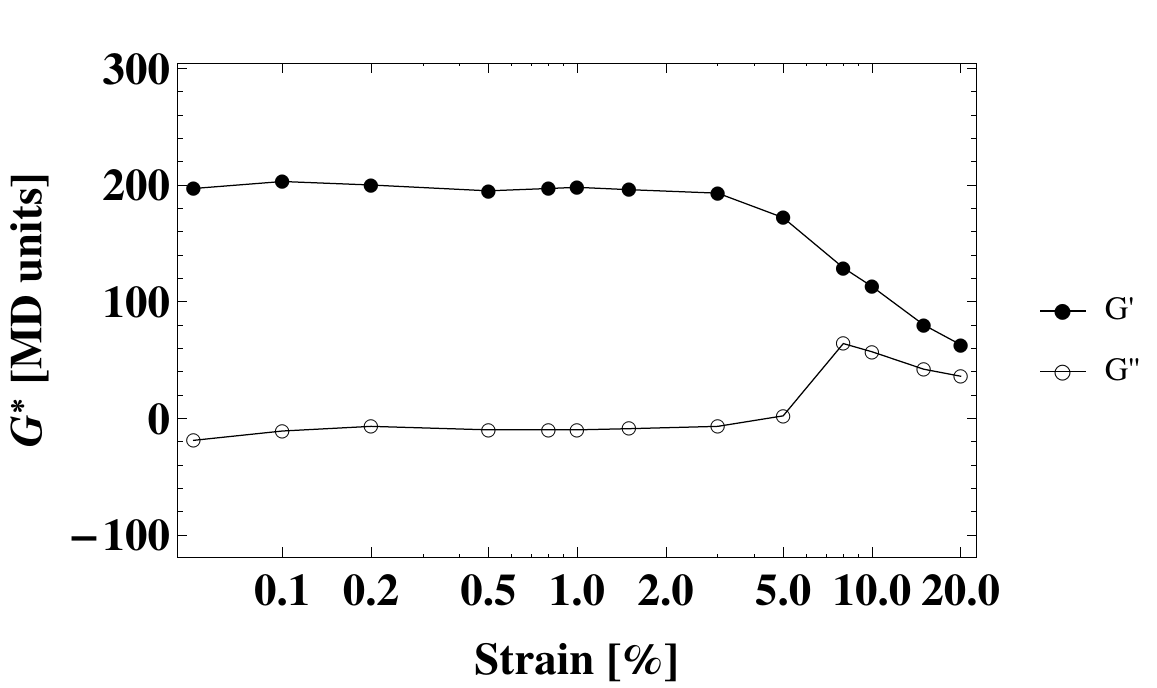}}
	\subfigure[$T=3, \phi/\phi^{\ast}=2.7$]{\includegraphics[width=0.45\textwidth]{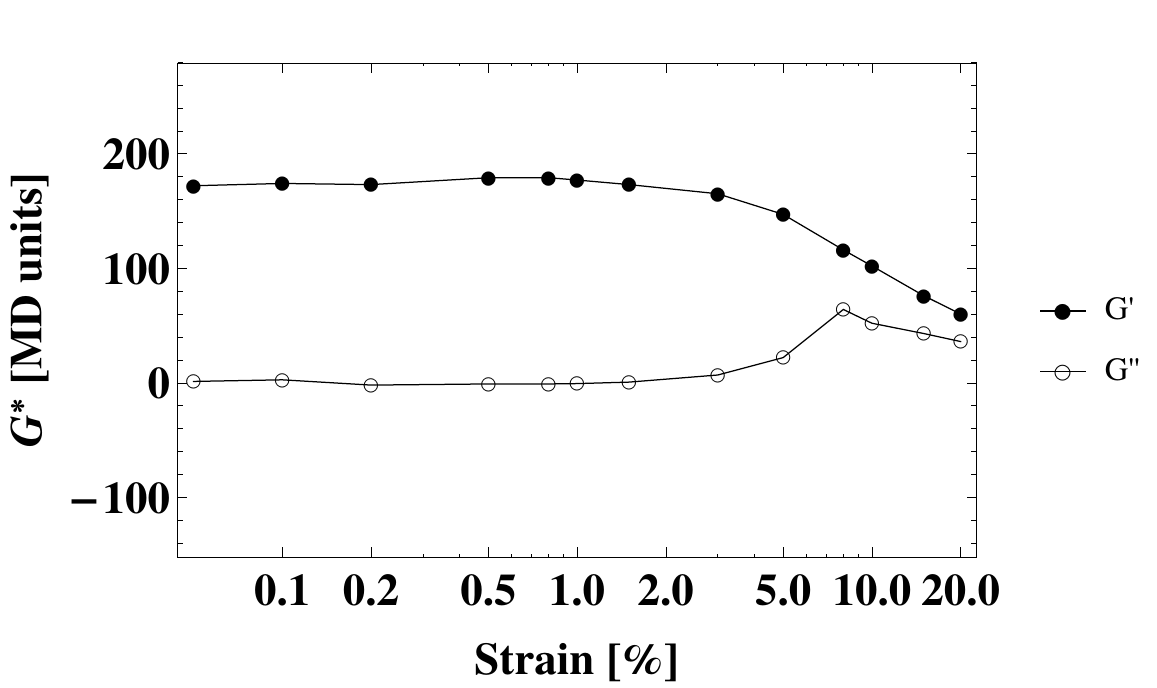}}
	\subfigure[$T=4, \phi/\phi^{\ast}=2.7$]{\includegraphics[width=0.45\textwidth]{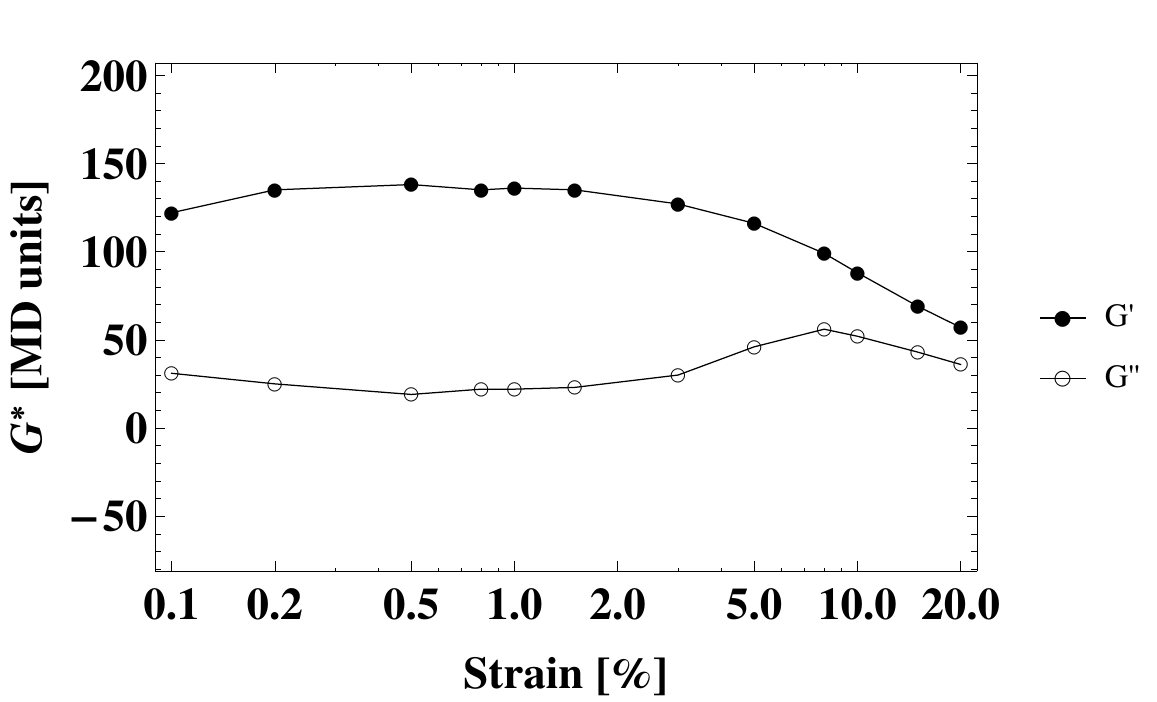}}
	\subfigure[$T=5,\phi/\phi^{\ast}=2.7$]{\includegraphics[width=0.45\textwidth]{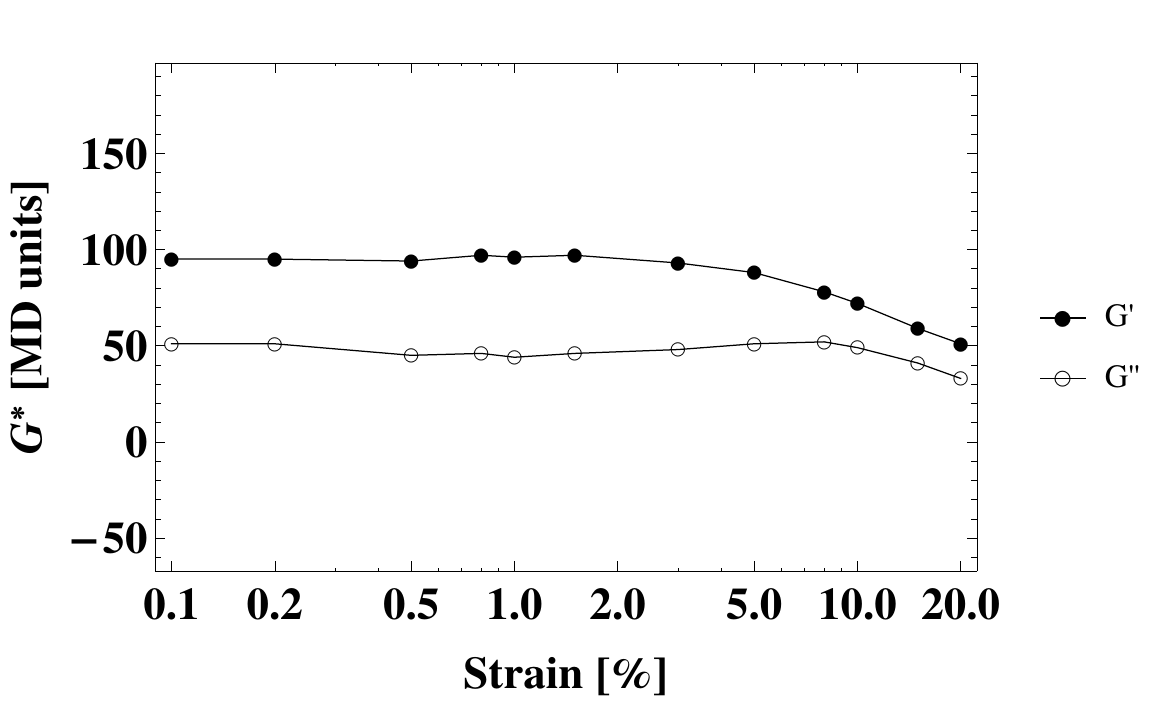}}
	\subfigure[$T=6, \phi/\phi^{\ast}=2.7$]{\includegraphics[width=0.45\textwidth]{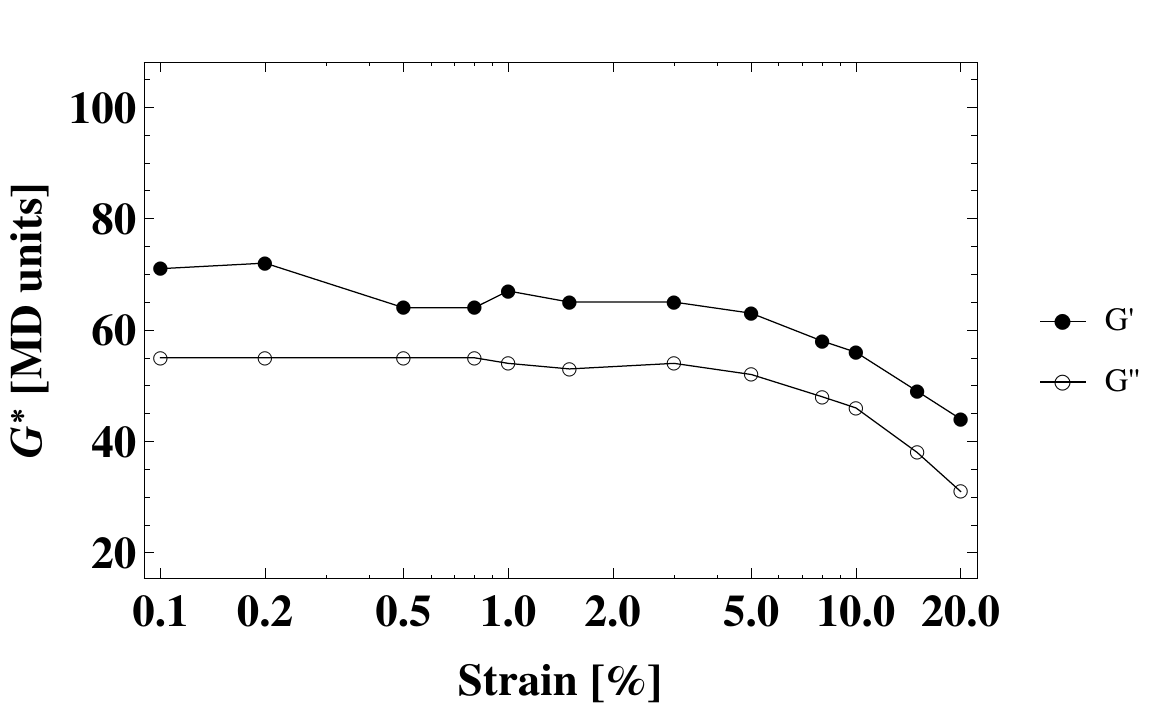}}
\caption{Dynamic moduli $G'$ and $G''$, measured in strain-controlled settings at $\omega=0.2$ and $\phi/\phi^{\ast}=2.7$ at various temperatures. $G'$ is shown by solid symbols and $G"$ is indicated by open symbols.}
\label{fig:lowrate temp}
\end{figure*}

\newpage

\twocolumngrid

\begin{figure*}
\centering
	\subfigure[ $T=1$]{\includegraphics[width=0.4\textwidth]{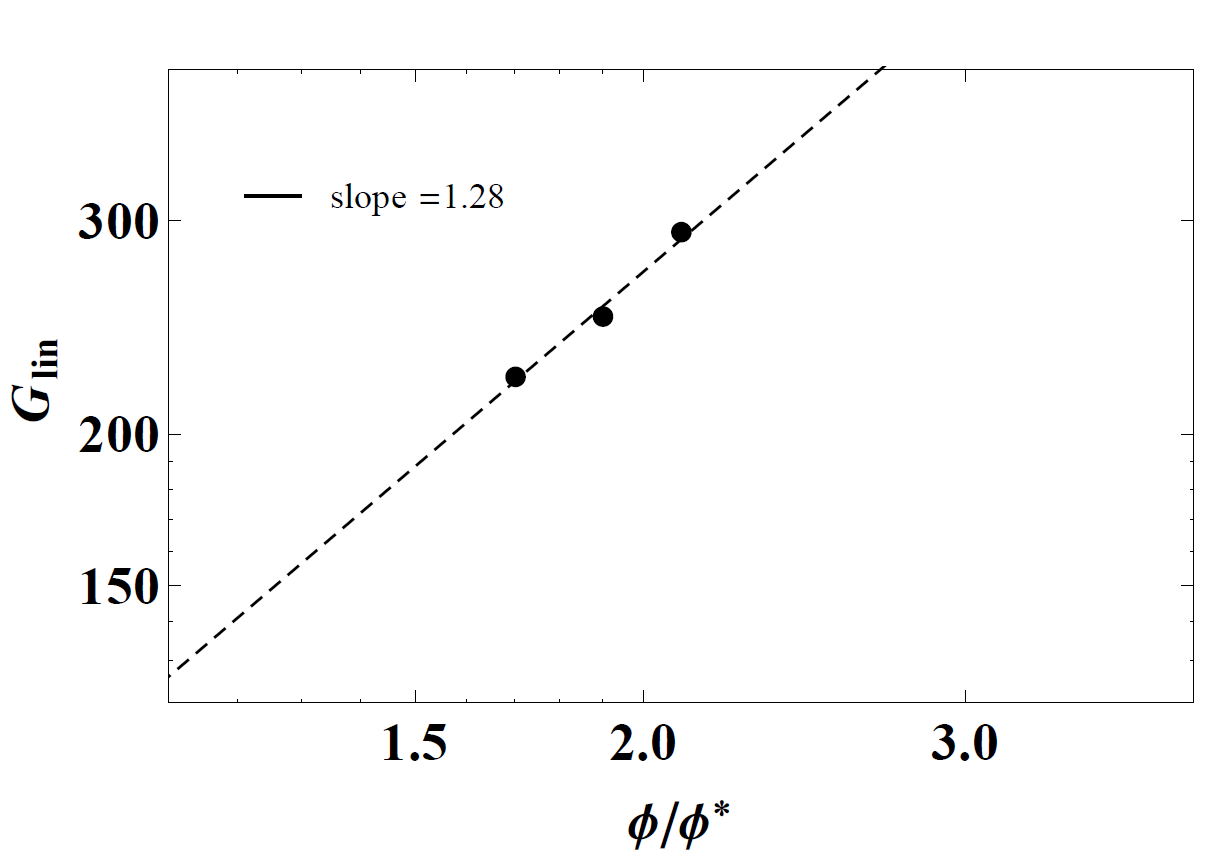}}
		\subfigure[ $\phi/\phi^{\ast}=2.7$]{\includegraphics[width=0.4\textwidth]{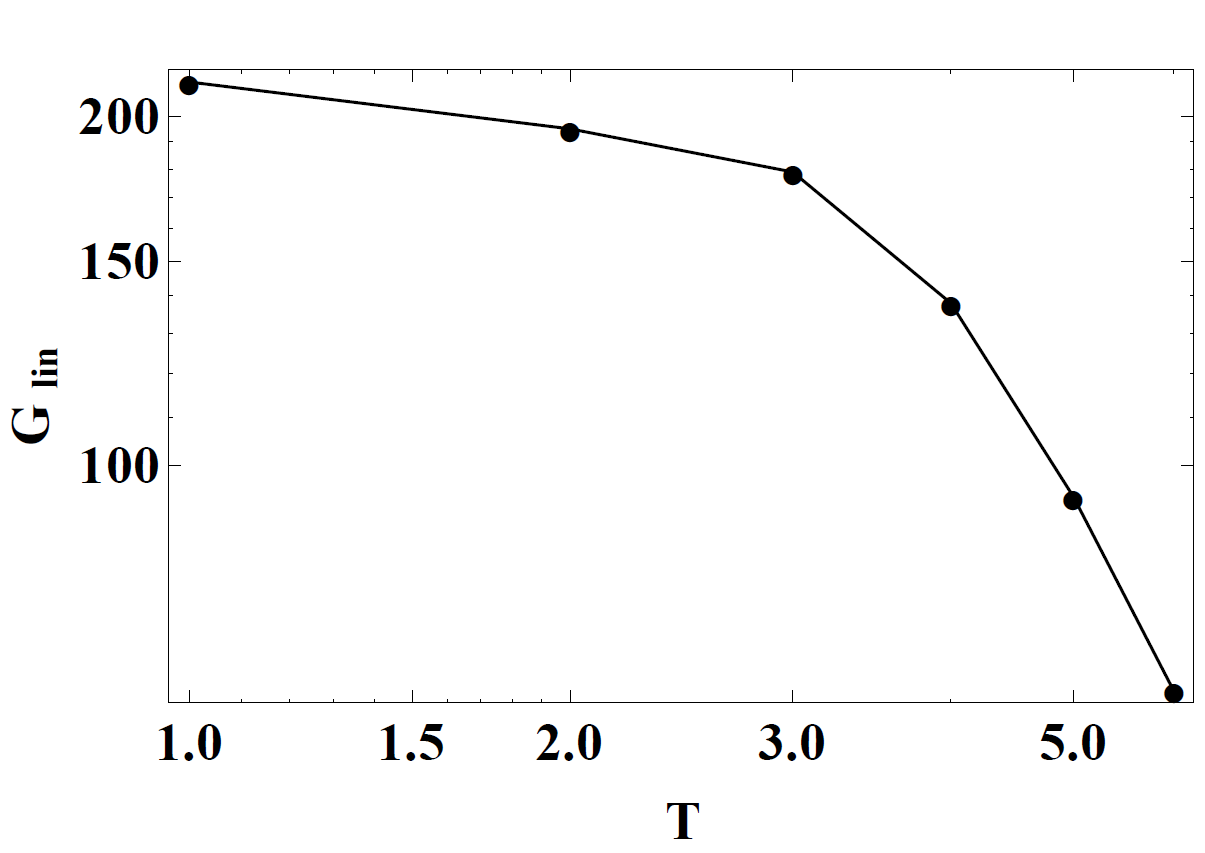}}
		
\caption{The linear modulus as a function of temperature and density for B$3$L$45$ at $\omega=0.2$. (a) The slope of power-law fit is $1.28$. }
\label{fig:glin}
\end{figure*}

The temperature of the simulation also influences the modulus of the system. Using Eq. (\ref{eq:temperature}), we can change the temperature in the simulation. It affects the mechanics in two ways: first, the self assembly process is affected as elevated temperatures favor high entropy states more than high energy states, which skews the architectures towards smaller cluster sizes. This tends to lower $G'$. Secondly, the mechanical response due to the linkers is itself temperature dependent: entropically elastic effects scale with temperature. This latter effect would raise the storage modulus as a function of temperature, as is the case in rubber elasticity. In the Fig. [\ref{fig:lowrate temp}] the strain dependent shear modulus for $\dot{\gamma} \leq 20\%$ at different temperatures is shown. The modulus is calculated based on the stress data from simulations that include all the contributions of the stress tensor. The result shows a linear regime up to $\gamma=5\%$, after which $G'$ drops off rapidly. Linear modulus as a function of  temperature and density is shown in a logarithmic plot in  Fig. [\ref{fig:glin}]. The slope of the power law fit to the temperature dependant linear modulus from numerical results is $1.28$, that is close to theoretical prediction for spherical flower-like micelles, i.e. 1.95. 

\begin{figure}[ht]\centering
  \includegraphics[width=.5\textwidth]{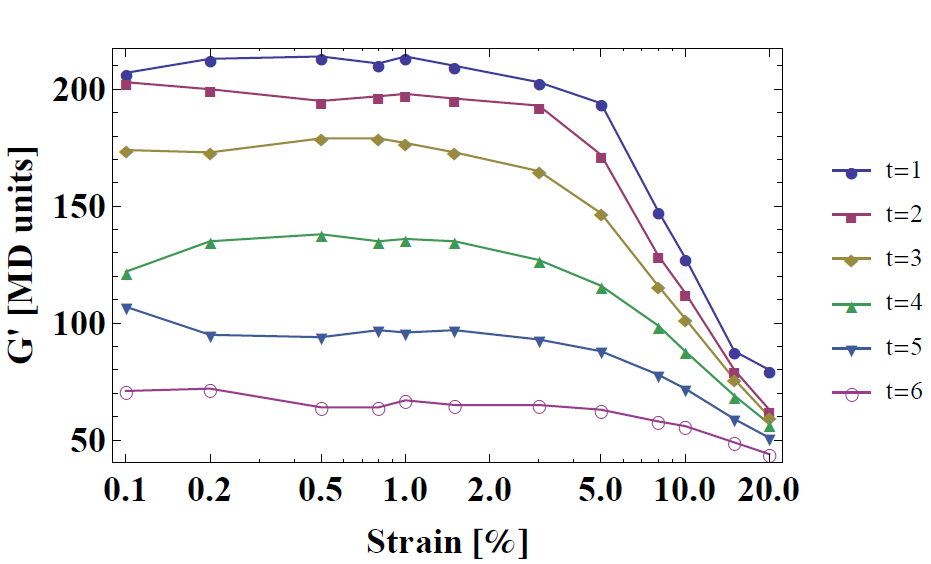}
  \caption{Dynamic moduli $G'$, measured in strain-controlled settings at $\omega=0.2$ and $ \phi/\phi^{\ast}=2.7$ at various temperatures. Different colors corresponds to different temperatures as indicated in the legend.}\label{fig:temp comp}
  
\end{figure}

The yield point for all of the systems in Fig. [\ref{fig:lowrate temp}] is where the $G'$ starts to drop, at $\gamma\approx 5\%$. That this is the yield point is also evidenced by the concurrent and characteristic rise in $G''$. To investigate the effect of the temperature on the modulus we plot the elastic modulus for all temperatures as a function of strain in \ref{fig:temp comp}. The experimental measurements at large strains gives a quantitative comparison of the contribution of the elastic response at different temperatures\citep{zfahimithesis}. A decrease in the magnitude of the modulus with the temperature is observed which is in good agreement with the simulation results we find here Fig. [\ref{fig:temp comp}]. The fact that the overall effect of temperature in these networks is to {\em lower} the modulus suggests that, apparently, the effect of temperature on the aggregate size is dominant over the single-chain elastic contribution. This suggests that rubber elasticity theories should be modified to include also the temperature dependent changes in connectivity to fully capture the mechanical response of the system (and that, therefore, the model presented in section \ref{sec2} may not offer a complete description.)

\section{Conclusion}
\label{sec8}

Our MD simulations suggest that the motif of using repeating hydrophilic and hydrophobic blocks gives rise to interesting self-assembling phase behavior. In a large regime of phase space (defined by a combination of $\ve$ and $N_l$) we observe, in NVT MD simulations, network phases where small clusters of the hydrophobic blocks are connected by (typically) single hydrophilic linker chains to other such clusters, giving rise to a "single molecule networks". The connectivity of this network is generally determined by a combination of the size of an aggregate, which sets the number of outgoing linkers and thus the potential for bridges, and the number of aggregates which sets the number of potential partners for such bridging. At intermediate values of $\ve$ and $N_l$, the combination between "supply and demand" of linkers is optimal which should result, in the highest moduli for the network material. The multiblock amphiphile system thus allows direct control over its supramolecular arrangement through molecular design, and thus to mechanical properties. This suggests much richer applications for these systems than what has been established thus far, and in particular makes them suitable candidates for exploring future biomimetic mechanical performance. 

\begin{acknowledgments}
We thank Hans Wyss, Rint Sijbesma, Gajanan Pawar, Zahra Fahimi, Marcel Koenigs, Christian Moerland, Thijs Michels, Alexey Lyulin and Arlette Baljon for enlightening discussions and for sharing experimental details and results. Funding from the Eindhoven University of Technology's High Potential programme is gratefully acknowledged.
\end{acknowledgments}

\bibliography{references_modeling_multiscale}

\end{document}